\documentclass[prb,preprint,onecolumn]{revtex4}%
\usepackage{amsfonts}
\usepackage{amsmath}
\usepackage{amssymb}
\usepackage{graphicx}%
\begin{document}
\title{Field induced superconducting phase in Superconductor-Normal metal and
Superconductor-Superconductor bilayer}
\author{X. M\textsc{ontiel}}
\affiliation{Condensed Matter Theory Group, LOMA, UMR 5798, Universit\'{e} Bordeaux 1,
33405 Talence France}
\author{A. I. B\textsc{uzdin}}
\altaffiliation{Also at Institut Universitaire de France, Paris.}

\affiliation{Condensed Matter Theory Group, LOMA, UMR 5798, Universit\'{e} Bordeaux 1,
33405 Talence France}


\begin{abstract}
We study the proximity effect in a superconductor (S)-normal metal (N) bilayer
systems under in-plane magnetic field and demonstrate that a compensation
between the Zeeman effect and the energy splitting between bonding and
anti-bonding levels may lead to a magnetic field induced superconducting phase
well above the standard paramagnetic limit. It occurs that the non-uniform
Fulde-Ferrell-Larkin-Ovchinikov superconducting state also exists in the field
induced phase. The presence of the impurities scattering shrink the region of
field induced superconductivity existence in S-N and S-S bilayers.

\end{abstract}
\date[Date text]{date}
\maketitle

\section{ INTRODUCTION}

Quasi two dimensional superconductors have been studied for fifty years. A
strong upper critical magnetic field $\left(  H_{c2}\right)  $ anisotropy was
observed for the first time in intercalated layered crystal of dichalcogenides
of transition metals \cite{bulaevskii11}. In these compounds, $H_{c2}$ is
higher for the in-plane orientation (ab plane) than in the perpendicular one
(c-axis). Moreover, this $H_{c2}$ anisotropy was observed in intercalated
graphite superconductors (in $C_{8}K$ \cite{Hannay}$^{,}$\cite{Koike} in
$C_{6}Ca$ \cite{Jobiliong}$^{,}$\cite{Weller} and in $C_{6}Yb$ \cite{Weller}),
in organic superconductors \cite{BuzdinBulaevskii}, in high superconducting
critical temperature (High $T_{c}$) cuprates superconductors \cite{Welp}$^{-}%
$\cite{Tuominen}) and also in superconducting-superconducting (S-S')
$YBa_{2}Cu_{3}O_{7}/DyBa_{2}Cu_{3}O_{7}$ and superconducting-insulating (S-I)
$\left(  YBa_{2}Cu_{3}O_{7}\right)  _{n}/\left(  \Pr Ba_{2}Cu_{3}O_{7}\right)
_{m}$ artificial superlattices \cite{Triscone1}$^{,}$\cite{Triscone 2}. High
$T_{c}$ cuprates superconductors have a layered crystal structure \cite{Cyrot}
and a strong electron anisotropy \cite{Welp}$^{,}$\cite{Cyrot}$^{-}%
$\cite{Nachtrab}. The superconducting coherence length along the c-axis
$\xi_{c}$ is smaller than the interlayer distance $d$. Consequently, high
$T_{c}$ cuprate superconductors can be considered as natural superlattices. In
high $T_{c}$ cuprate superconducting compounds, superconductivity exists in
$CuO_{2}$ atomic planes which are sandwiched by non-superconducting atomic
planes \cite{Tinkham}$^{,}$\cite{Cyrot}.

Ginzburg-Landau model (in the weak anisotropy limit $\left(  \xi_{c}\lesssim
d\right)  $) \cite{VL Ginzburg} and Lawrence-Doniach model (in the strong
anisotropy limit $\left(  \xi_{c}\ll d\right)  $) \cite{Lowrence doniach} give
the description of the $H_{c2}$ anisotropic properties in layered
superconductors near $T_{c}$. This $H_{c2}$ anisotropy in superconducting
multilayers can also be described microscopically by the standard
Bardeen-Cooper-Schrieffer (BCS) and the tunneling Hamiltonian theory. Using
this method, \ we obtain the $\left(  H_{c2},T\right)  $ phase diagram of
layered superconducting systems.

Some of high $T_{c}$ cuprate superconductors can be considered as a stack of
S-N, S-S' \cite{Buzdin2} or S-F \cite{Andreev1} weakly coupled bilayers. The
S-N, S-S' or S-F \ bilayer constitute the elemental unit cell of the
multilayer. The properties of the S-N, S-S' or S-F bilayers qualitatively
differs from a single S, N or F layers. Consequently, the properties of
multilayers based on single layer elemental unit cell may be qualitatively
different of the properties of multilayers based on bilayer elemental unit
cell. We show in this paper that $\left(  H,T\right)  $ phase diagram, with in
plane magnetic field, of \ S-N and S-S' bilayers may reveal a magnetic field
induced superconducting phase.

\begin{figure}[ptb]
\begin{center}
\includegraphics[scale=0.3,angle=0]{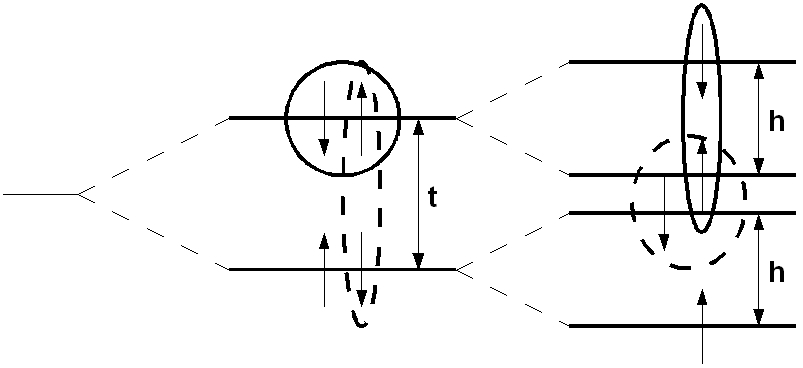}
\end{center}
\caption{Mechanism of compensation of the Zeeman effect by the degeneracy
lifting between the bonding and anti-bonding state.}%
\label{Fig_compensation}%
\end{figure}

\begin{figure}[ptb]
\begin{center}
\includegraphics[scale=0.3,angle=0]{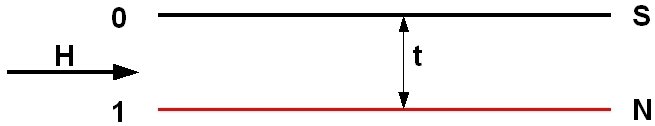}
\end{center}
\caption{A superconducting (S)-Normal metal (N) bilayer with in-plane magnetic
field H.}%
\label{SN1}%
\end{figure}

The case of S-F multilayers has been studied in \cite{Andreev1}$^{-}%
$\cite{Prokic}. In S-S bilayer, Buzdin \textit{et al} have demonstrated
\cite{Buzdin 21}$^{,}$\cite{Buzdin22} the possibility to overcome the
paramagnetic limit at low temperature for a high in-plane critical magnetic
field. The field induced superconducting phase may appear at high magnetic
field if the interlayer coupling energy $t$ is higher than $T_{c}$. Moreover,
in this phase, the adjacent S layers have\ opposite signs of the order
parameter (this is so-called $\pi$-state \cite{Andreev1}). In this case, the
Zeeman effect is compensated by the bonding-antibonding degeneracy lift
produced by the hybridization between the two S layers (see figure
$\ref{Fig_compensation}$). The Cooper pairs in the $\pi$ state are more stable
at high magnetic field than the $0$-state. The $0$-state occurs when the
adjacent S layers support the same signs of the order parameter. Somewhat
similar idea in the context of two-band superconductivity was introduced by
Kulic and Hofmann \cite{Kulic}. In this paper, we show that in a S-N bilayer at high in-plane
magnetic field $H$ (see figure $\ref{SN1}$), at low temperature and strong
enough coupling $t>T_{c0}$ between the two planes, the paramagnetic limit is
also enhanced above the usual Fulde-Ferrell-Larkin-Ovchinikov (FFLO)
\cite{LO}$^{,}$\cite{FF} limit and a field induced superconducting phase may
appear at high magnetic field. The corresponding mechanism is qualitatively
the same as in the S-S bilayer but naturally there is no $\pi$ state
realization in this case. We study also the influence of the impurity
scattering on the $\left(  H,T\right)  $ phase diagram of S-N and S-S bilayers.

The outline of the paper is as following. In Sec. II, we present the model of
a multilayer system and give the exact solutions of the Eilenberger equations.
In Sec. III, we study the influence of the transfer integral on the
superconducting critical temperature and the effect of impurities. In section
IV, we investigate the phase diagram of the S/N bilayer in the strong exchange
field regime in both the clean and the dirty limits. In section V, we study
the influence of impurities of the S-S bilayer $\left(  H,T_{c}\right)  $
phase diagram.

\section{MODEL OF AN ATOMIC THICKNESS S/N BILAYER}

We start with a non-interacting model (see for example \cite{Bulaevskii1}%
$^{,}$\cite{Andreev1}) of layered systems with alternating superconducting and
normal metal layers. The electron motion is described in the N and S layers by
the spin-dependent energy spectra $\xi_{n,\sigma}\left(  \mathbf{k}\right)  $
and $\xi_{s,\sigma}\left(  \mathbf{k}\right)  $ respectively. The parameters
that characterize the systems are the transfer energy between the N and S
layers $t$, the Cooper pairing constant $\lambda$ which is assumed to be
nonzero in S layers only. The Zeeman energy splitting, due to in plane
magnetic field $H$, is written as $h=\mu_{B}H$ where $\mu_{B}$ is the Bohr magneton.

The two mechanisms destroying superconductivity under a magnetic field are the
orbital and the paramagnetic effect \cite{De Gennes}$^{,}$\cite{D St James et
Sarma}. Usually it is the orbital effect that is more restrictive. However, in
systems with a large effective mass of electrons \cite{Bianchi}$^{,}%
$\cite{Capan} or in low-dimensional compounds, like quasi-one dimensional or
layered superconductors under in-plane magnetic field \cite{Uji}, the orbital
magnetism is weakened and it is the paramagnetic effect that is responsible
for superconductivity destruction.

The Chandrasekhar-Clogston paramagnetic limit \cite{Chandrasekar}$^{,}%
$\cite{Clogston} is achieved when the polarization energy of the normal
electron gas, $\chi_{n}H^{2}/2$, equals the superconducting condensation
energy $N\left(  0\right)  \Delta_{0}^{2}/2$, where $N\left(  0\right)  $ is
the density of state of the normal electron gas, $\chi_{n}$ is its spin
susceptibility, and $\Delta_{0}=1.76T_{c}$ is the zero temperature
superconducting gap. This criterion gives the exchange field $h_{p}\left(
T=0\right)  =\Delta_{0}/\sqrt{2}$ where the superconductor should undergo a
first-order transition to the normal state. Larkin and Ovchinikov \cite{LO}
and Fulde and Ferrell \cite{FF} (FFLO) predicted the existence of a
non-uniform superconducting state with slightly higher critical field
$h_{FFLO}^{3D}\left(  T=0\right)  =0.755\Delta_{0}>h_{p}\left(  T=0\right)  $.
For quasi-2D superconductors the critical field of the FFLO state is even
higher, namely $h_{FFLO}^{2D}\left(  T=0\right)  =\Delta_{0}$,
\cite{Bulaevskii} while in quasi one dimensional systems there is no
paramagnetic limit at all \cite{Buzdin1}. We focus on the 2D case for which a
generic temperature magnetic field phase diagram has been established
\cite{Bulaevskii}.

We consider the case when the coupling between the layers is realized via the
transfer energy $t$. In the whole paper, we assume $t\ll E_{F}$ \ where
$E_{F}$ is the Fermi energy and then Cooper pairs are localized within each
plane. The layers are coupled together by the coupling Hamiltonian
\begin{equation}
\hat{H}_{t}=t\sum_{j,\sigma,\mathbf{k}}\left[  \psi_{j+1,\sigma}^{+}\left(
\mathbf{k}\right)  \psi_{j,\sigma}\left(  \mathbf{k}\right)  +\psi_{j,\sigma
}^{+}\left(  \mathbf{k}\right)  \psi_{j+1,\sigma}\left(  \mathbf{k}\right)
+H.c\right]  . \label{hamiltonien_tunnel}%
\end{equation}

where $\psi_{j,\sigma}^{+}\left(  \mathbf{k}\right)  $ (resp.$~\psi_{j,\sigma
}\left(  \mathbf{k}\right)  $) is the creation (resp. annihilation) operator
of an electron with spin $\sigma$ and momentum $\mathbf{k}$ in the $j$th
layer. In this paper, we study the S-N and S-S bilayers. In S-N system, the
superconducting layer has the index $j=0$ \ and the normal metal $j=1$. In the
S-S bilayer, the superconducting layers are indexed $j=0$ and $j=1$. The
Hamiltonian of the system can be written as :%
\begin{equation}
\hat{H}=\hat{H}_{0}+\hat{H}_{BCS}+\hat{H}_{t}, \label{hamiltonian_general}%
\end{equation}
where $H_{0}$ is the kinetic and Zeeman Hamiltonian , $H_{t}$ the tunneling
Hamiltonian and $H_{BCS}$ the BCS Hamiltionian. For the $j$th layer, the
kinetic and Zeeman parts of the Hamiltonian are written as
\begin{equation}
\hat{H}_{0}=\sum_{\sigma,\mathbf{k}}\left[  \xi_{j,\sigma}\left(
\mathbf{k},h_{j}\right)  \psi_{j,\sigma}^{+}\left(  \mathbf{k}\right)
\psi_{j,\sigma}\left(  \mathbf{k}\right)  \right]  ,
\label{hamiltonien_zeeman_kinetic}%
\end{equation}
The Zeeman effect manifests itself in breaking the spin degeneracy of the
electronic energy levels according to
\begin{equation}
\xi_{j,\sigma}\left(  \mathbf{k},h_{j}\right)  =\xi_{j}\left(  \mathbf{k}%
\right)  -\sigma h_{j}%
\end{equation}
where $\xi_{j}\left(  \mathbf{k}\right)  =\mathbf{k}^{2}/2m-E_{F}$ i.e. for
simplicity we choose the same electron spectrum in both layers.

The field $h_{j}$ in the $j$th layer is assumed to be the same in both layers
$\left(  h_{0}=h_{1}=h\right)  $. We suppose an s-wave singlet
superconductivity coupling which is treated in $H_{BCS}$ within a mean field
approximation \cite{b.abrikosov_gorkov}%
\begin{equation}
\hat{H}_{BCS}=\sum_{j,\mathbf{k}}\left[  \Delta_{j}^{\ast}\left(
\mathbf{q}\right)  \psi_{j,\downarrow}^{+}\left(  \mathbf{k}\right)
\psi_{j,\uparrow}^{+}\left(  -\mathbf{k}\right)  +\Delta_{j}\left(
\mathbf{q}\right)  \psi_{j,\uparrow}\left(  \mathbf{k}\right)  \psi
_{j,\downarrow}\left(  -\mathbf{k}\right)  \right]  +\frac{1}{\left\vert
\lambda\right\vert }\int\mathbf{d}^{2}\mathbf{r}\Delta_{j}^{2}\left(
\mathbf{r}\right)  \label{hamiltonien_BCS}%
\end{equation}
where $\boldsymbol{r}$ is the two-dimensional coordinate within each layer and
$\lambda$ the electron-electron coupling constant in the S layer only. The
superconducting order parameter $\Delta_{j}$ is non zero only in the S layers
as the coupling constant is $0$ in the N layer. In order to investigate the
occurrence of modulated superconducting phase (FFLO), we choose the
superconducting order parameter in the form
\[
\Delta\left(  \mathbf{r}\right)  =\Delta e^{i\mathbf{q}.\mathbf{r}}%
\]
where $\mathbf{q}$ is the FFLO modulation wave vector. Using Gorkov's
formalism, we introduce the normal $G$ and anomalous $\widetilde{F}$ Green
functions \cite{b.abrikosov_gorkov} :%
\begin{equation}%
\begin{array}
[c]{c}%
G_{j,l}\left(  \mathbf{k},\mathbf{k}^{\prime}\right)  =-\left\langle T_{\tau
}\left(  \psi_{\uparrow,j}\left(  \mathbf{k}\right)  \psi_{\uparrow,l}%
^{+}\left(  \mathbf{k}^{\prime}\right)  \right)  \right\rangle =\delta\left(
\mathbf{k}-\mathbf{k}^{\prime}+\mathbf{q}\right)  G_{j,l}\left(
\mathbf{k}\right)  ,\\
F_{j,l}^{+}\left(  \mathbf{k},\mathbf{k}^{\prime}\right)  =\left\langle
T_{\tau}\left(  \psi_{\downarrow,j}^{+}\left(  \mathbf{k}\right)
\psi_{\uparrow,l}^{+}\left(  \mathbf{k}^{\prime}\right)  \right)
\right\rangle =\delta\left(  \mathbf{k}+\mathbf{k}^{\prime}\right)
F_{j,l}^{+}\left(  \mathbf{k}\right)  ,
\end{array}
\label{fonction_green_general}%
\end{equation}
where the brackets mean statistical averaging over grand-canonical
distribution and $T_{\tau}$ the ordering operator in the Matsubara's formalism
\cite{b.abrikosov_gorkov}, and $j$ and $l$ the layer's indexes. From the
equation of motion \cite{b.abrikosov_gorkov}, the system of Green functions
equation is in the Fourier representation \ in the S-N bilayer:%
\[
\left(
\begin{array}
[c]{cccc}%
\left(  i\omega-\xi_{0,\uparrow}\left(  \mathbf{k+q}\right)  \right)  & -t &
\Delta_{0} & 0\\
-t & \left(  i\omega-\xi_{1,\uparrow}\left(  \mathbf{k+q}\right)  \right)  &
0 & 0\\
\Delta_{0}^{\ast} & 0 & \left(  i\omega+\xi_{0,\downarrow}\left(
\mathbf{k}\right)  \right)  & t\\
0 & 0 & t & \left(  i\omega+\xi_{1,\downarrow}\left(  \mathbf{k}\right)
\right)
\end{array}
\right)  .\left(
\begin{array}
[c]{c}%
G_{0,0}\left(  \mathbf{k+q}\right) \\
G_{1,0}\left(  \mathbf{k+q}\right) \\
F_{0,0}^{+}\left(  \omega,\mathbf{k}\right) \\
F_{1,0}^{+}\left(  \omega,\mathbf{k}\right)
\end{array}
\right)  =\left(
\begin{array}
[c]{c}%
1\\
0\\
0\\
0
\end{array}
\right)  ,
\]

where $\omega=\left(  2n+1\right)  \pi T$ are the fermionic Matsubara
frequencies. In quasi 2D superconductors, the maximal modulus of the FFLO wave
vector is of the order of $\left(  \xi_{0}\right)  ^{-1}$, $\xi_{0}$ being the
typical superconducting coherence length. Since $\xi_{0}\gg\frac{1}{k_{F}}$
which of the order of the inter atomic distance with a good approximation we
can consider $\xi_{j,\uparrow}\left(  \mathbf{k}+\mathbf{q}\right)
=\xi\left(  \mathbf{k}\right)  -h+\mathbf{v}_{F}.\mathbf{q}$ where
$\mathbf{v}_{F}$ is the Fermi velocity vector in the plane. The anomalous
Green function in the S layer writes%
\[
F_{0,0}^{+}=\frac{-\Delta_{0}^{\ast}A}{-\alpha_{0}A-\beta t^{2}+t^{4}}%
\]
where $A=\left(  i\omega-\xi\left(  \mathbf{k}\right)  +h-v_{F}.q\right)
\left(  i\omega+\xi\left(  \mathbf{k}\right)  +h\right)  $, $\alpha
_{0}=\left\vert \Delta_{0}\right\vert ^{2}-\left(  i\omega-\xi\left(
\mathbf{k}\right)  +h-v_{F}.q\right)  \left(  i\omega+\xi\left(
\mathbf{k}\right)  +h\right)  $ and $\beta=\left(  i\omega-\xi\left(
\mathbf{k}\right)  +h-v_{F}.q\right)  ^{2}+\left(  i\omega+\xi\left(
\mathbf{k}\right)  +h\right)  ^{2}$.

The superconducting order parameter in the $0$th superconducting layer
satisfies the self-consistency equation
\begin{equation}
\Delta_{0}^{\ast}=\left\vert \lambda\right\vert T\sum_{\omega>0}%
\sum_{\mathbf{k}}F_{0,0}^{+}=\left\vert \lambda\right\vert T\sum_{\omega}%
\int_{-\infty}^{+\infty}F_{0,0}^{+}d\xi. \label{Self_consistency}%
\end{equation}

To describe the FFLO modulated phase and the influence of the impurities it is
more convenient to use the quasi-classical Eilenberger formalism. Moreover, we
include the FFLO modulation phase and non-magnetic impurities. Applying
Eilenberger's method \cite{Eilenberger} for layered system \cite{Kopnin} with
Hamiltonian $\left(  \ref{hamiltonian_general}\right)  $, the system of
equations of Green functions can be written as:%

\begin{equation}
\left(
\begin{array}
[c]{cccc}%
\widetilde{\omega}-i\mathbf{v}_{F}.\mathbf{q} & -i\frac{t}{2} & 0 & i\frac
{t}{2}\\
-i\frac{t}{2} & \widetilde{\omega}-i\mathbf{v}_{F}.\mathbf{q} & i\frac{t}{2} &
0\\
0 & i\frac{t}{2} & \widetilde{\omega}-i\mathbf{v}_{F}.\mathbf{q} & -i\frac
{t}{2}\\
i\frac{t}{2} & 0 & -i\frac{t}{2} & \widetilde{\omega}-i\mathbf{v}%
_{F}.\mathbf{q}%
\end{array}
\right)  .\left(
\begin{array}
[c]{c}%
f_{0,0}^{+}\\
f_{1,0}^{+}\\
f_{1,1}^{+}\\
f_{0,1}^{+}%
\end{array}
\right)  =\left(
\begin{array}
[c]{c}%
\Delta_{0}^{\ast}+\frac{\left\langle f_{0,0}^{+}\left(  \omega,\mathbf{q}%
\right)  \right\rangle _{\phi}}{2\tau}\\
0\\
0\\
0
\end{array}
\right)  \label{equation_Eilenberger_SN}%
\end{equation}
where $\widetilde{\omega}=\omega+ih+\left(  1/2\tau\right)  $ and $f_{j,l}%
^{+}\left(  \omega,\mathbf{q}\right)  =\frac{1}{i\pi}\int_{-\infty}^{+\infty
}d\xi F_{j,l}^{+}\left(  \omega,\xi,\mathbf{q}\right)  d\xi$ is the anomalous
Green function in the Eilenberger formalism and $\tau$ electron mean free pass
time. We write $\mathbf{v}_{F}.\mathbf{q}=v_{F}.q.\cos\left(  \phi\right)  $
where $\phi$ is the polar angle $\left(  \mathbf{v}_{F},\mathbf{q}\right)  $
and $\left\langle {}\right\rangle _{\phi}$ is the average over $\phi$. We
assume an in-plane scattering on impurities and the absence of spin flip
during the electron-impurity interaction. To consider the presence of
impurities we substitute $\omega$ by $\omega+1/2\tau$ and $\Delta_{j}^{\ast}$
by $\Delta_{j}^{\ast}+\left\langle f_{j,j}^{+}\left(  \omega,\mathbf{q}%
\right)  \right\rangle _{\phi}/2\tau$ see for example \cite{Kopnin}.

Solving the Eilenberger equation $\left(  \ref{equation_Eilenberger_SN}%
\right)  $ yields the Eilenberger Green function for the S layer labeled
$j=0$
\begin{equation}
f_{0,0}^{+}=\frac{\Delta_{0}^{\ast}}{2\left\{  1-\frac{\left(  \frac{1}{2\tau
}\right)  \left[  \Omega_{1}\Omega_{3}+\Omega_{2}\Omega_{3}+2\Omega_{1}%
\Omega_{2}\right]  }{4\Omega_{1}\Omega_{2}\Omega_{3}}\right\}  }\left\{
\frac{1}{\omega_{3}}+\frac{1}{2\omega_{1}}+\frac{1}{2\omega_{2}}\right\}
\label{f_SN}%
\end{equation}
where we pose $\Omega_{1,2}^{2}=\left(  \widetilde{\omega}_{\pm}^{2}%
+v^{2}q^{2}\right)  $, $\Omega_{3}^{2}=\left(  \widetilde{\omega}^{2}%
+v^{2}q^{2}\right)  $ with $\widetilde{\omega}=\omega+ih+\left(
1/2\tau\right)  $, $\widetilde{\omega}_{\pm}=\omega+ih+\left(  1/2\tau\right)
\pm it$, $\omega_{3}=\widetilde{\omega}-iv_{F}.q.\cos\left(  \phi\right)  $
with $\left(  \omega_{1,2}=\omega_{3}\pm it=\widetilde{\omega}_{\pm}%
-iv_{F}.q.\cos\left(  \phi\right)  \right)  $. The averaged solution on the
$\phi$ angle of $\left(  \ref{f_SN}\right)  $ writes
\begin{equation}
\left\langle f_{0,0}^{+}\right\rangle _{\phi}=\frac{\Delta_{0}^{\ast}%
}{2\left\{  1-\frac{\left(  \frac{1}{2\tau}\right)  \left[  \Omega_{1}%
\Omega_{3}+\Omega_{2}\Omega_{3}+2\Omega_{1}\Omega_{2}\right]  }{4\Omega
_{1}\Omega_{2}\Omega_{3}}\right\}  }\left\{  \frac{1}{\Omega_{3}}+\frac
{1}{2\Omega_{1}}+\frac{1}{2\Omega_{2}}\right\}  \label{f_general_SN}%
\end{equation}
where $\left\langle {}\right\rangle _{\phi}$ is the average on the $\phi$
angle. Close to the superconducting critical temperature of the second order
phase transition, the self consistency $\left(  \ref{Self_consistency}\right)
$ can be written \cite{D St James et Sarma}
\begin{equation}
\ln\left(  \frac{T_{c}}{T_{c_{0}}}\right)  =\operatorname{Re}\left(
\sum_{\omega>0}\left(  \left\langle \widetilde{f}_{\downarrow\uparrow}%
^{0,0}\left(  \omega,q\right)  \right\rangle _{\phi}-\frac{\pi}{\omega
}\right)  \right)  \label{Self_consistency_1}%
\end{equation}
where $T_{c}$ is the critical temperature of the superconducting layer in the
S-N bilayer and $T_{c0}=\frac{2\gamma\omega_{D}}{\pi}e^{-\frac{2\pi^{2}%
}{\left\vert \lambda\right\vert mk_{F}}}$ the critical temperature of an
isolated superconducting layer with $m$ the electron's mass, $k_{F}$ the Fermi
impulsion, $\gamma=0.577215$ is the Euler's constant and $\omega_{D}$ the
Debye frequency.

At zero temperature, close\ to the critical magnetic field of the second order
phase $h_{0}$, the order parameters $\Delta_{j}$ are also small the self
consistency $\left(  \ref{Self_consistency}\right)  $ writes
\begin{equation}
\ln\left(  \frac{h}{h_{0}}\right)  =\frac{2T_{c}}{\pi}\int_{0}^{+\infty
}\operatorname{Re}\left(  \left\langle \widetilde{f}_{\downarrow\uparrow
}^{0,0}\left(  \omega,q\right)  \right\rangle _{\phi}-\frac{\pi}{\omega
+ih_{0}}\right)  d\omega. \label{Self_consistency_2}%
\end{equation}

\section{PROXIMITY\ EFFECT\ IN\ S-N BILAYER}

In this section, we investigate the superconducting phase in the S layer in
the clean limit $\left(  \tau\rightarrow\infty\right)  $. We study the
superconducting critical temperature as a function of the interlayer coupling.
\ We obtain the critical magnetic field of second order superconducting \ to
normal metal phase transition as a function of the temperature and the
interlayer coupling. Study of the influence of the impurities and of the S-S
bilayers are respectively proposed in section IV and V.

\subsection{Critical temperature}

We study first the influence of the proximity effect on the superconducting
critical temperature $T_{c}$ of the $S$ layer when no magnetic field is
applied $\left(  h=q=0\right)  $ in the clean limit $\left(  \tau
\rightarrow\infty\right)  $. Then $\left(  \ref{f_general_SN}\right)  $
becomes :%
\begin{equation}
f_{0,0}^{+}=\tfrac{\Delta^{\ast}\left(  t^{2}+2\omega^{2}\right)  }%
{2\omega\left(  t^{2}+\omega^{2}\right)  },\label{formule h0 1}%
\end{equation}
thus the self consistency equation \ writes%
\[
\ln\left(  \frac{T_{c}}{T_{c0}}\right)  =-\frac{1}{4}\left[  2\gamma
+4\ln2+\Psi\left(  \frac{1}{2}-\frac{it}{2\pi T_{c}}\right)  +\Psi\left(
\frac{1}{2}+\frac{it}{2\pi T_{c}}\right)  \right]  .
\]
\begin{figure}[ptb]
\begin{center}
\includegraphics[scale=0.3,angle=0]{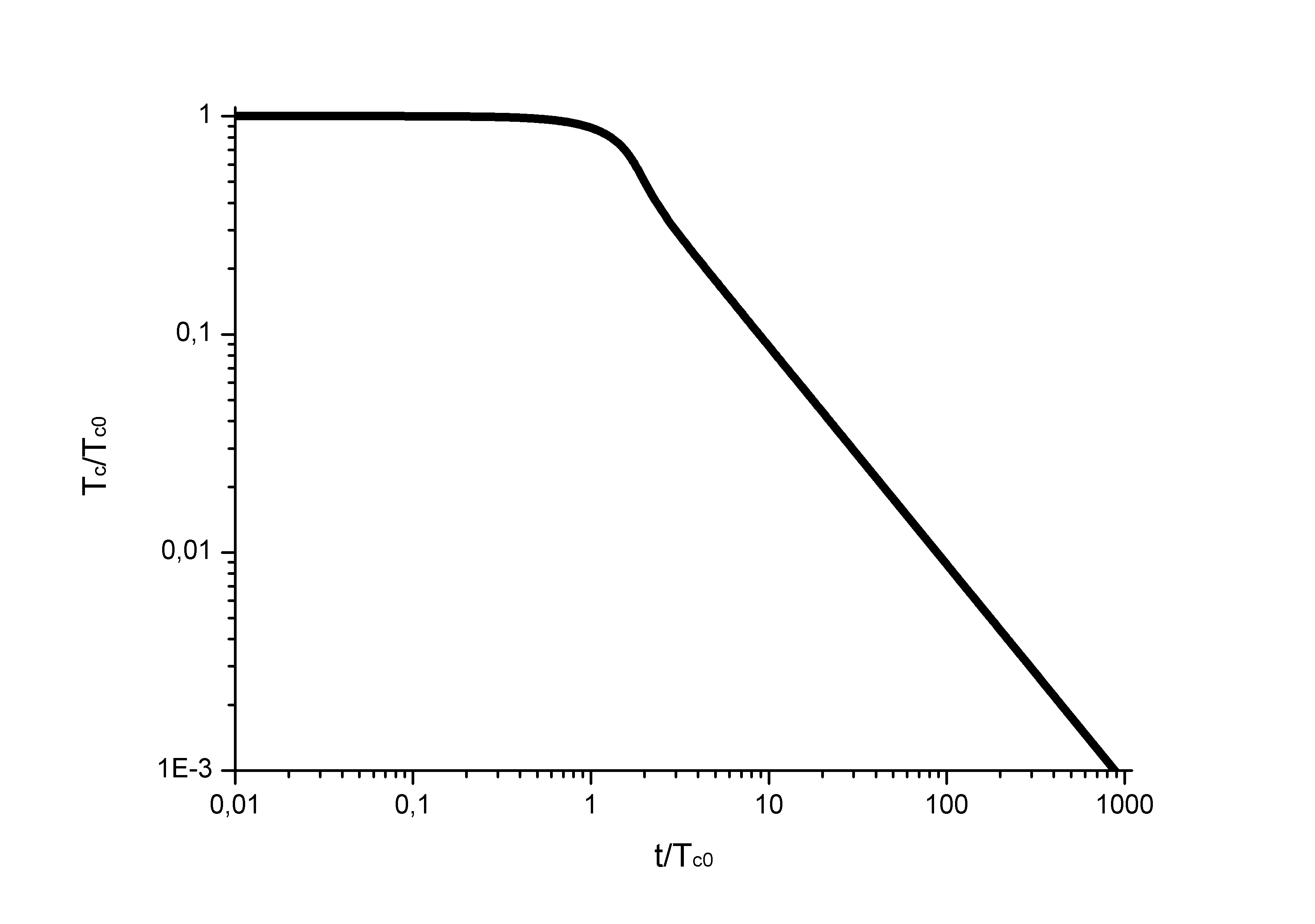}
\end{center}
\caption{Graph of $T_{c}/T_{c0}$ as a function of $t/T_{c0}$ (solid line). For
$t\gg T_{c0}$, the critical temperature of the superconductor decreases to
zero.}%
\label{FIG_1}%
\end{figure}where $\Psi\left(  x\right)  $ is the digamma function. As seen in
fig $\ref{FIG_1}$, the superconducting critical temperature decreases with the
increase of the proximity effect. At low transfer energy $t\ll T_{c}$, the
superconducting critical temperature varies like $\frac{T_{c}}{T_{c0}}%
=1-\frac{7}{8}\frac{\zeta\left(  3\right)  }{\pi^{2}}\left(  \frac{t}{T_{c0}%
}\right)  ^{2}$. In the case of low interlayer coupling, the superconducting
critical temperature reveals a quadratic decrease with the transfer energy.
The superconducting state is not qualitatively influenced by the normal metal
layer and can be considered as a single superconducting layer.

At strong coupling between S and N layers at $t\gg T_{c}$ (but in the limit
$t\ll\omega_{D}$) , the superconducting temperature varies as $\frac{T_{c}%
}{T_{c0}}=\frac{\pi e^{-\gamma}}{2}\frac{T_{c0}}{t}$. The critical temperature
decreases with the tunneling transfer as more and more Cooper pairs leak into
the N layer. The superconducting properties in the N and S layers are
practically the same and the bilayer can be considered as an equivalent single
S layer with an effective coupling constant $\widetilde{\lambda}$ where
$\widetilde{\lambda}<\lambda$. In the case where $t\gg\omega_{D}$, the S-N
bilayer can be considered as a single superconducting layer S with
$\widetilde{\lambda}=\frac{\left\vert _{\lambda}\right\vert }{2}$ as predicted
in \cite{Bulaevskii1}.

\subsection{Phase diagram of the S-N bilayer}

We study the $\left(  h,T\right)  $ and $\left(  h,t\right)  $ phase diagram
of the S-N bilayer in the clean case and in presence of non-magnetic
impurities. In a two-dimensional S monolayer, we can define three critical
magnetic fields at zero temperature. $h_{0}=\Delta_{0}/2$ is the critical
magnetic field for a second order phase transition. $h^{I}=\Delta_{0}/\sqrt
{2}$ is the critical magnetic field for a first order phase transition defined
by Clogston-Chandrasekar \cite{Chandrasekar}$^{,}$\cite{Clogston}.
$h^{FFLO}=\Delta_{0}$ is the critical magnetic field in the presence of FFLO
modulations. One can see that $h^{FFLO}>h^{I}>h_{0}$. In a clean S monolayer
with an applied in-plane magnetic field, the critical field is $h^{FFLO}$
,\cite{D St James et Sarma}.

In this case, the Eilenberger anomalous Green function $\left(  \ref{f_SN}%
\right)  $ becomes for arbitrary interlayer coupling $t$ :%
\begin{equation}
f^{+}\left(  0,0\right)  =\tfrac{\Delta^{\ast}}{2}\left[  \tfrac{1}{\left(
\omega+ih+i\overrightarrow{v}_{F}.\overrightarrow{q}\right)  }+\tfrac
{1}{2\left(  \omega-it+ih+i\overrightarrow{v}_{F}.\overrightarrow{q}\right)
}+\tfrac{1}{2\left(  \omega+it+ih+i\overrightarrow{v}_{F}.\overrightarrow
{q}\right)  }\right]  \label{fsnssimp}%
\end{equation}
where we note the appearance of three energy scales $E_{3}=h+\overrightarrow
{v}_{F}.\overrightarrow{q}$ and $E_{1,2}=h\pm t+\overrightarrow{v}%
_{F}.\overrightarrow{q}$.

\subsubsection{$\left(  h,t\right)  $ phase diagram at zero
temperature\bigskip}

\begin{figure}[ptb]
\begin{center}
\includegraphics[scale=0.3,angle=0]{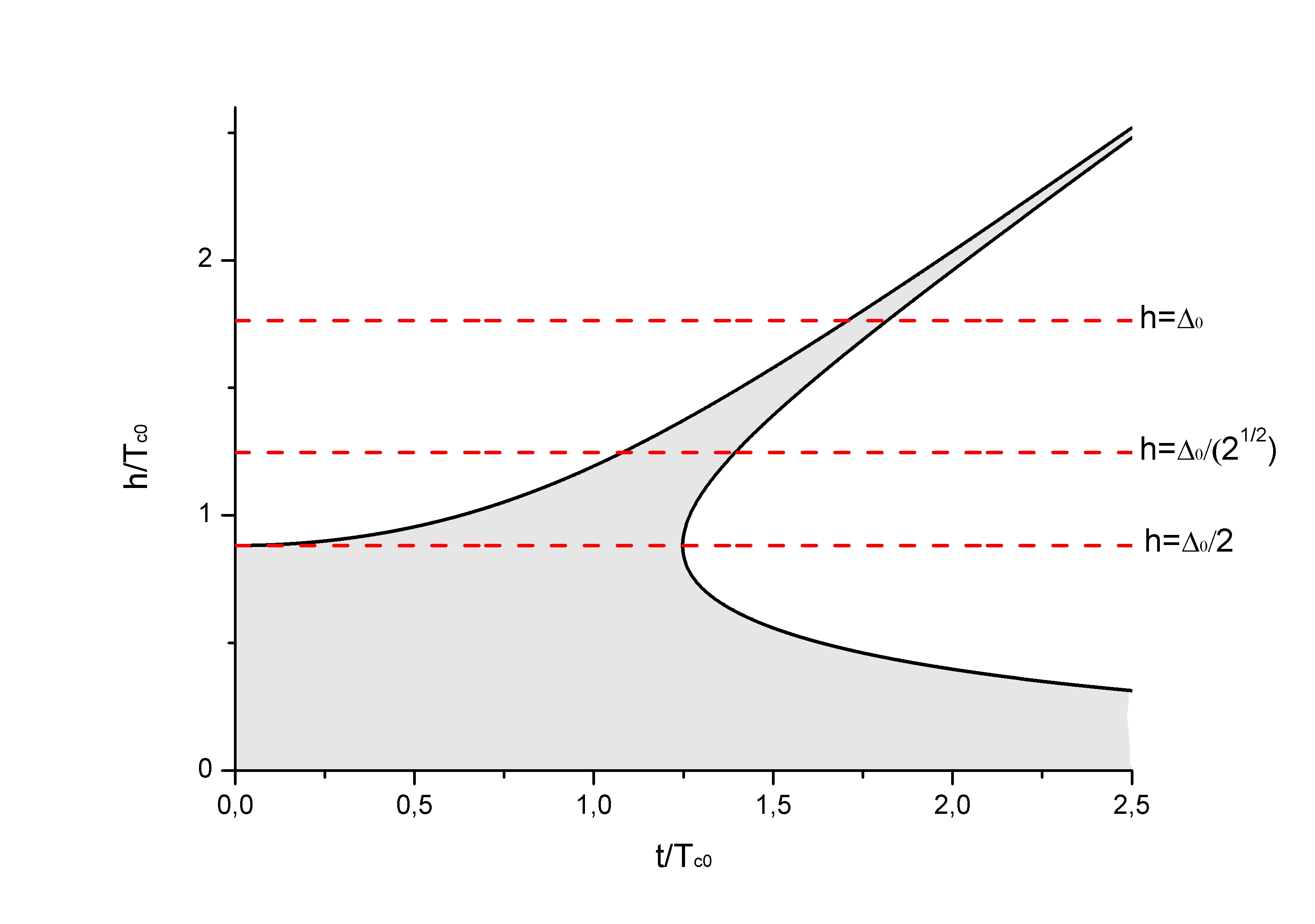}
\end{center}
\caption{$\left(  h/T_{c0},t/T_{c0}\right)  $ diagram for the S-N bilayer in
the clean limit $\left(  \tau\rightarrow\infty\right)  $at T=0K (solid line).
The uniform superconducting state is presented in grey. The line $h=\Delta
_{0}/2$ presents the critical magnetic field for a second order
superconducting phase transition for a single superconducting layer. The line
$h=\Delta_{0}$ corresponds the FFLO paramagnetic limit for a single
superconducting layer. The line $h=\Delta_{0}/\sqrt{2}$ represents the first
order paramagnetic limit for a single superconducting layer.}%
\label{Fig_hfcttSNC0}%
\end{figure}

From $\left(  \ref{fsnssimp}\right)  $and the self consistency equation
$\left(  \ref{Self_consistency_2}\right)  $ , the critical magnetic field $h$
is shown to satisfy%
\begin{equation}
\left\vert h_{c}-t+\sqrt{\left\vert \left(  h_{c}-t\right)  ^{2}-\left(
q.v_{F}\right)  ^{2}\right\vert }\right\vert .\left\vert h_{c}+t+\sqrt
{\left\vert \left(  h_{c}+t\right)  ^{2}-\left(  q.v_{F}\right)
^{2}\right\vert }\right\vert .\left\vert h_{c}+\sqrt{\left\vert h_{c}%
^{2}-\left(  q.v_{F}\right)  ^{2}\right\vert }\right\vert ^{2}=h_{0}^{4}
\label{eq_self_con_2FFLOT0}%
\end{equation}
where one must find the value of $q$ that maximizes the critical field $h_{c}%
$. If the field induced phase is assumed to be uniform in the each planes,
namely if $q=0$, equation $\left(  \ref{eq_self_con_2FFLOT0}\right)  $ merely
reduces to
\begin{equation}
\left\vert h_{c}\right\vert ^{2}.\left\vert h_{c}-t\right\vert .\left\vert
h_{c}+t\right\vert =h_{0}^{4} \label{eq_self_con_2FFLOT01}%
\end{equation}
The number of solutions with physical meaning of the equations $\left(
\ref{eq_self_con_2FFLOT01}\right)  $ differs with the value of $t$ (see figure
$\ref{Fig_hfcttSNC0}$ ) . We defines the critical interlayer coupling
$t_{c}=\sqrt{2}h_{0}=1.2473T_{c0}$ that determines the number of physical solutions.

If $t<t_{c}$, the equation $\left(  \ref{eq_self_con_2FFLOT01}\right)  $ has
only one solution. The critical magnetic field at zero temperature writes
$h_{c1}=\frac{1}{2}\sqrt{2t^{2}+2\sqrt{t^{4}+4h_{0}^{4}}}$. In the limit $t\ll
T_{c0}$, the solution can be written $h_{c1}=\frac{\Delta_{0}}{2}\left(
1+\frac{t^{2}}{\Delta_{0}^{2}}\right)  $. We note that the critical magnetic
field at $T=0K$ in the S-N bilayer increases with the interlayer coupling $t$.

In the case $t>t_{c}$, the equation $\left(  \ref{eq_self_con_2FFLOT01}%
\right)  $ has three solutions with physical meaning. The first solution is
$h_{c1}$. The second and the third solution are $h_{c2}=\frac{1}{2}%
\sqrt{2t^{2}+2\sqrt{t^{4}-4h_{0}^{4}}}$ and $h_{c3}=\frac{1}{2}\sqrt
{2t^{2}-2\sqrt{t^{4}-4h_{0}^{4}}}$ respectively. In the limit $t\gg T_{c0}$,
the three solutions can be written as $h_{c1,2}=t\pm\Delta_{0}^{4}/32t^{3}$
and $h_{c3}=\Delta_{0}^{2}/4t$. In the limit $t\gg T_{c0}$, $T_{c}$ is of
order of $T_{c0}^{2}/t$ and then $h_{c3}$ is of order of $T_{c}$.
Consequently, $h_{c3}$ define the lowest critical magnetic field.

For $t=t_{c}$ the critical fields $h_{c2}$ and $h_{c3}$ coincide.

In the case of high interlayer coupling $t>t_{c}$ , a field induced
superconducting phase appears at high magnetic field. This phase exists
between the two magnetic fields $h_{c1,2}=t\pm\Delta_{0}^{4}/32t^{3}$. Thus,
the new zero temperature paramagnetic limit $h_{c1}=t+\Delta\left(  0\right)
^{4}/32t^{3}$ \ may be tuned far above the usual one $h^{FFLO}=\Delta_{0}$
merely by increasing the interlayer coupling.

\begin{figure}[ptb]
\begin{center}
\includegraphics[scale=0.3,angle=0]{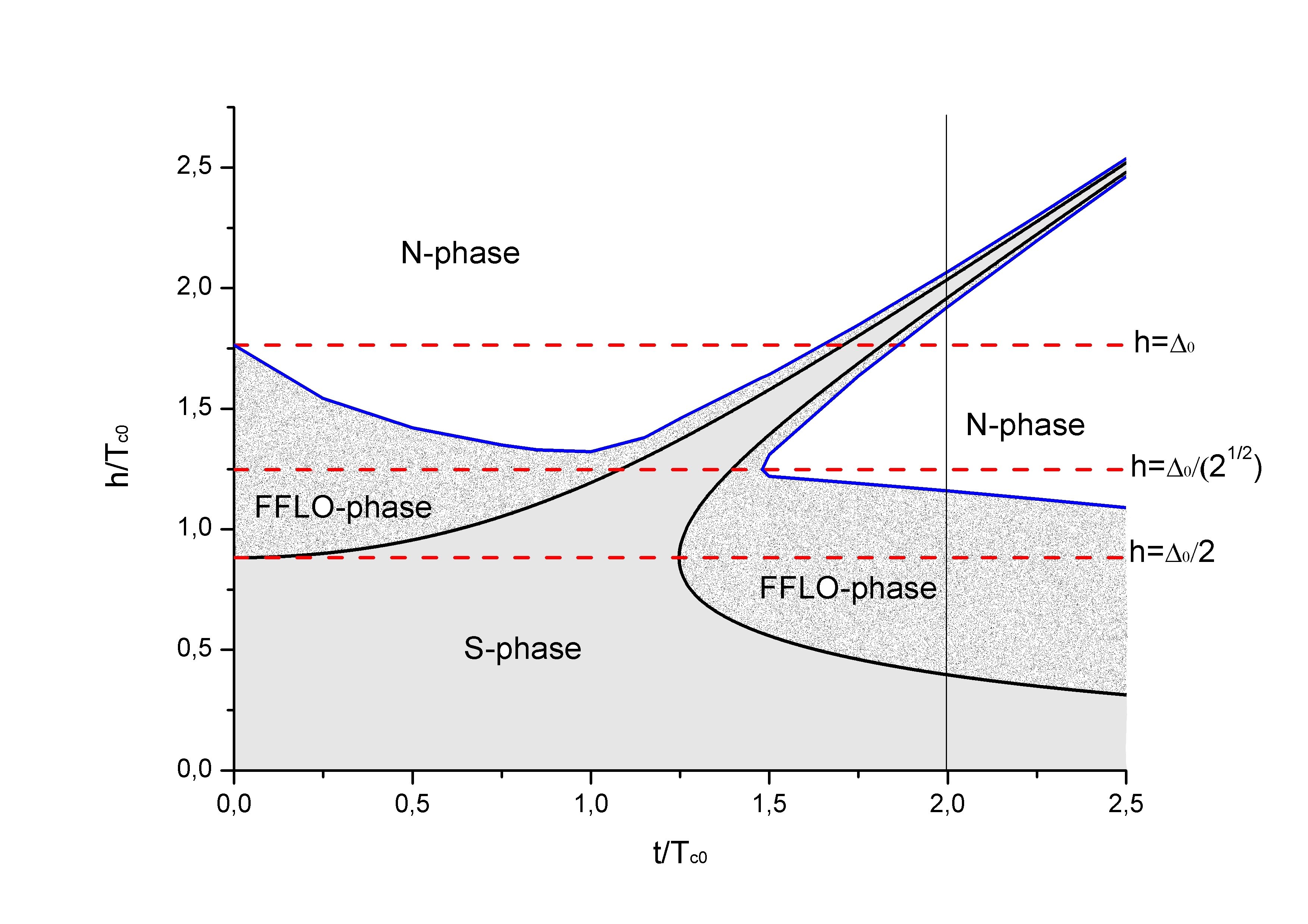}
\end{center}
\caption{$\left(  h/T_{c0},t/T_{c0}\right)  $ diagram for the S-N bilayer in
the clean limit $\left(  \tau\rightarrow\infty\right)  $ (solid line). The
uniform superconducting state is presented in the grey region. The non-uniform
superconducting (FFLO) phase in the S-N bilayer is presented in the dotted
region. The line $h=\Delta_{0}/2$ presents the critical magnetic field for a
second order superconducting phase transition for a single superconducting
layer. The line $h=\Delta_{0}$ represents the FFLO paramagnetic limit for a
single superconducting layer. The line $h=\Delta_{0}/\sqrt{2}$ represents the
first order paramagnetic limit for a single superconducting layer. }%
\label{Fig_hfcttSNC0FFLO}%
\end{figure}

Thorough analysis of equation $\left(  \ref{eq_self_con_2FFLOT0}\right)  $
shows that the upper critical field is even increased by an in-plane
modulation (see figure $\ref{Fig_hfcttSNC0FFLO}$).

The FFLO paramagnetic limit of the S-N bilayer also depends on the
interlayer coupling $t$ as seen in the figure $\ref{Fig_hfcttSNC0FFLO}$. The
field induced superconducting phase is observable at $T=0K$, only when
$h_{c2}$ and $h_{c3}$ are distinguishable. In presence of FFLO modulation, the
critical magnetic field at zero temperature $h_{c2}^{FFLO}$ and $h_{c3}%
^{FFLO}$ are separated in the case $t\gtrsim1.5T_{c0}$. Below this value, the
usual superconducting $\left(  h,T\right)  $ phase diagram may be strongly
deformed (see figure $\ref{FIG_h_fct_de_T_W2_FFLO1}$).

\subsubsection{$\left(  h,T\right)  $ phase diagram}

In this section, we study the second order $\left(  h,T\right)  $ phase
transition diagram taking into account FFLO modulation. The self consistency
equation $\left(  \ref{Self_consistency_1}\right)  $ is%
\begin{equation}%
\begin{array}
[c]{c}%
\ln\left(  \tfrac{T_{c0}}{T_{c}}\right)  =2T_{c}\int_{0}^{2\pi}\tfrac{d\phi
}{2\pi}\operatorname{Re}\left[  \gamma+2\ln\left(  2\right)  +\tfrac{1}{4}%
\Psi\left(  \tfrac{1}{2}+\tfrac{i\left(  h+v_{F}.q.\cos\left(  \phi\right)
\right)  }{2\pi T}\right)  \right]  +\\
+2T_{c}\int_{0}^{2\pi}\tfrac{d\phi}{2\pi}\operatorname{Re}\left[  \tfrac{1}%
{8}\Psi\left(  \tfrac{1}{2}+\tfrac{i\left(  h+t+v_{F}.q.\cos\left(
\phi\right)  \right)  }{\pi T}\right)  +\tfrac{1}{8}\Psi\left(  \tfrac{1}%
{2}+\tfrac{i\left(  h-t+v_{F}.q.\cos\left(  \phi\right)  \right)  }{\pi
T}\right)  \right]
\end{array}
.\label{eq_self_con_2FFLOT}%
\end{equation}
This analysis in general case can be performed only numerically on the basis
of the equation $\left(  \ref{eq_self_con_2FFLOT}\right)  $.
\begin{figure}[ptb]
\begin{center}
\includegraphics[scale=0.3,angle=0]{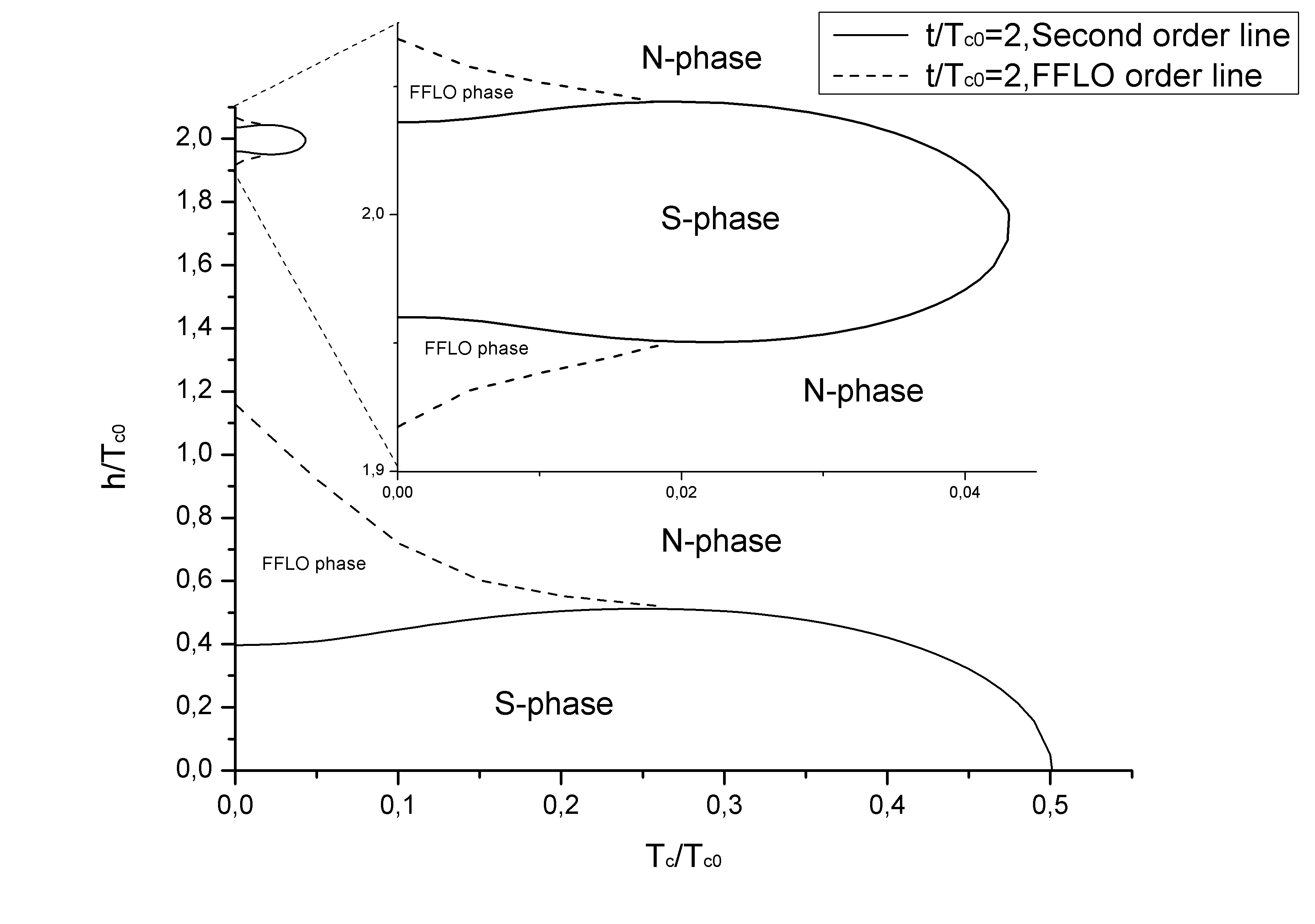}
\end{center}
\caption{$\left(  h/T_{c0},T_{c}/T_{c0}\right)  $ phase transition diagram
calculated for $t=2T_{c0}$ with the second order transition line (solid line)
and \ FFLO state to normal state transition line (doted line). The inset
presents a zoom of the superconducting re-entrance phase around $h\simeq
t\simeq2T_{c0}$.}%
\label{FIG_h_fct_de_T_W2_FFLO}%
\end{figure}\begin{figure}[ptbptb]
\begin{center}
\includegraphics[scale=0.3,angle=0]{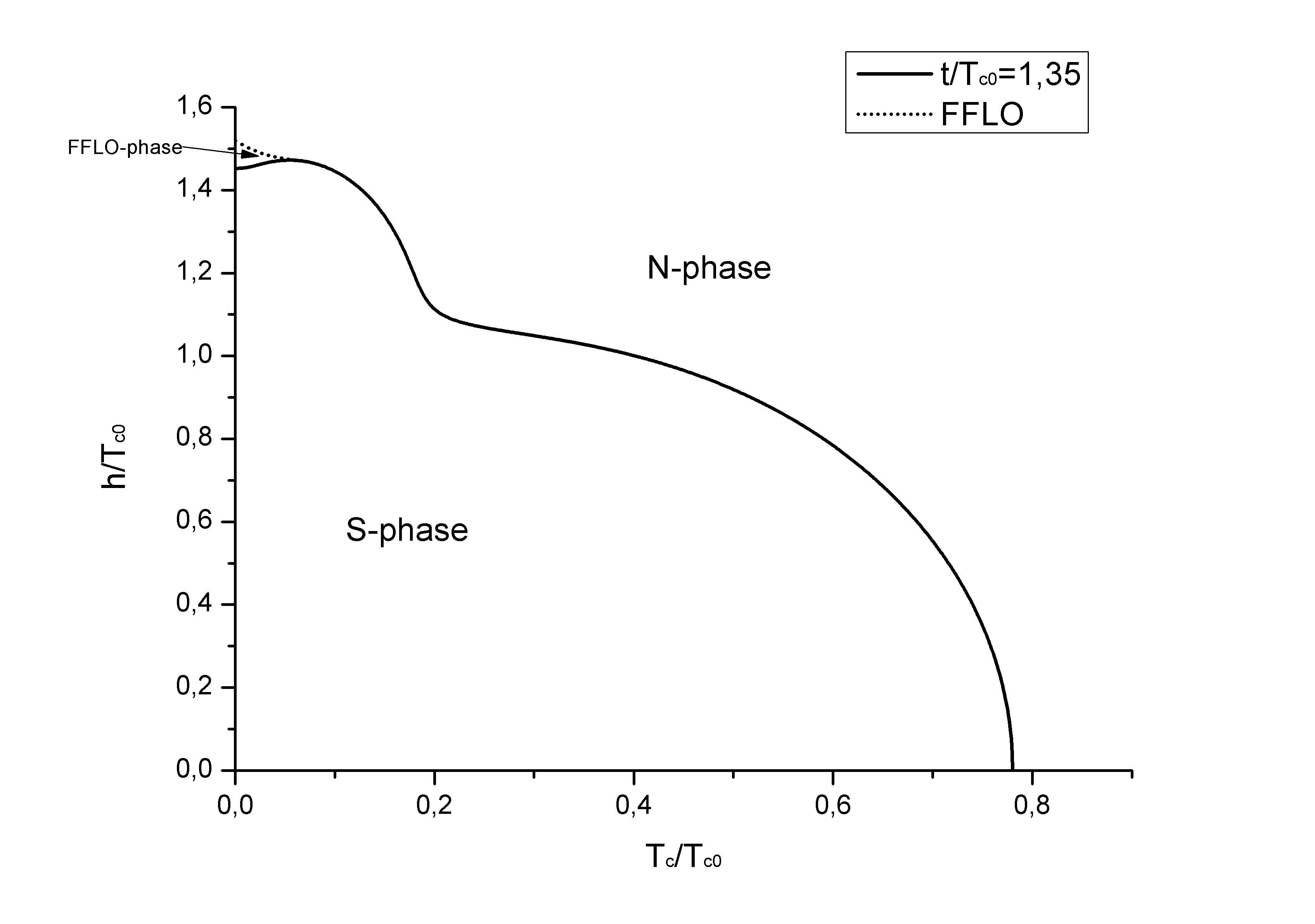}
\end{center}
\caption{$\left(  h/T_{c0},T_{c}/T_{c0}\right)  $ phase transition diagram
calculated for $t=1.35T_{c0}$ with the second order transition line (solid
line) and \ FFLO state to normal state transition line (doted line). We see
below $T_{c}\simeq0.2T_{c0}$ that the transition line is deformed. The
compensation between the Zeeman effect and the bonding and antibonding state
becomes relevant at low temperature.}%
\label{FIG_h_fct_de_T_W2_FFLO1}%
\end{figure}\begin{figure}[ptbptbptb]
\begin{center}
\includegraphics[scale=0.3,angle=0]{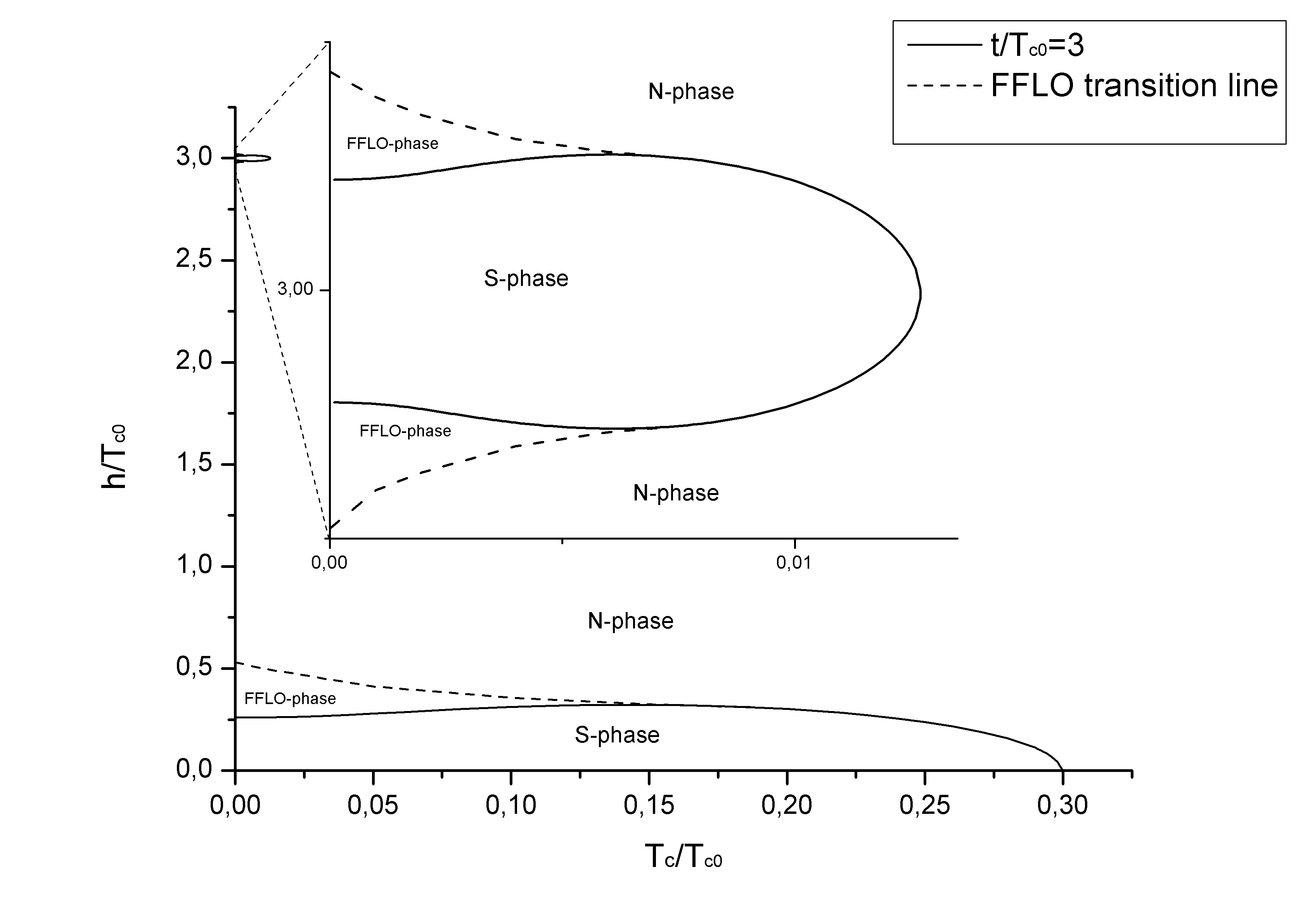}
\end{center}
\caption{$\left(  h/T_{c0},T_{c}/T_{c0}\right)  $ phase transition diagram
calculated for $t=3T_{c0}$ with the second order transition line (solid line)
and \ FFLO state to normal state transition line (doted line). The inset
presents a zoom of the superconducting re-entrance phase around $h\simeq
t\simeq3T_{c0}$.}%
\label{FIG_h_fct_de_T_W2_FFLO3}%
\end{figure}

A magnetic field induced superconducting state appears at high magnetic field
as we can see in the figure $\ref{FIG_h_fct_de_T_W2_FFLO}$ for $t=2.T_{c0}$
and $\ref{FIG_h_fct_de_T_W2_FFLO3}$ for $t=3.T_{c0}$. For $h\simeq t$, the
Zeeman effect that destroys the superconductivity is compensated by the
bonding-antibonding states degeneracy created by the proximity effect between
the S and the N layers (see figure $\ref{Fig_compensation}$). The lower and
upper critical lines merge at field $h=t$ and the field induced
superconductivity is confined to temperature lower than $T_{M}=\pi e^{-\gamma
}T_{c0}^{2}/\left(  8t\right)  $ in the limit $t\gg T_{c0}$. Therefore, the
superconducting field induced phase is confined to temperature lower than
\ $T_{M}$. These results \ were obtained for relatively strong coupling. For
lower coupling, $\left(  t\simeq T_{c0}\right)  $, the usual phase transition
diagram is strongly deformed as shown in figure $\ref{FIG_h_fct_de_T_W2_FFLO1}%
$ and finally disappear for $t$ smaller enough than $T_{c0}$. From an
experimental point of view, one might choose a system with an intermediate
coupling $t$ small enough to settle superconducting field induced phase but
large enough to separate re-entrance and usual S phase.

\section{EFFECT\ OF\ THE\ IMPURITIES ON\ THE
FIELD\ INDUCED\ SUPERCONDUCTING\ PHASE}

In this section, we investigate phases with uniform superconductivity in the S
layer. We study the superconducting critical temperature as a function of the
interlayer coupling. \ We obtain the critical magnetic field of second order
superconducting \ to normal metal phase transition as a function of the
temperature and the interlayer coupling. Study of the influence of the
impurities and of the S-S bilayers are respectively proposed in section IV and V.

\subsection{Critical temperature}

\begin{figure}[ptb]
\begin{center}
\includegraphics[scale=0.3,angle=0]{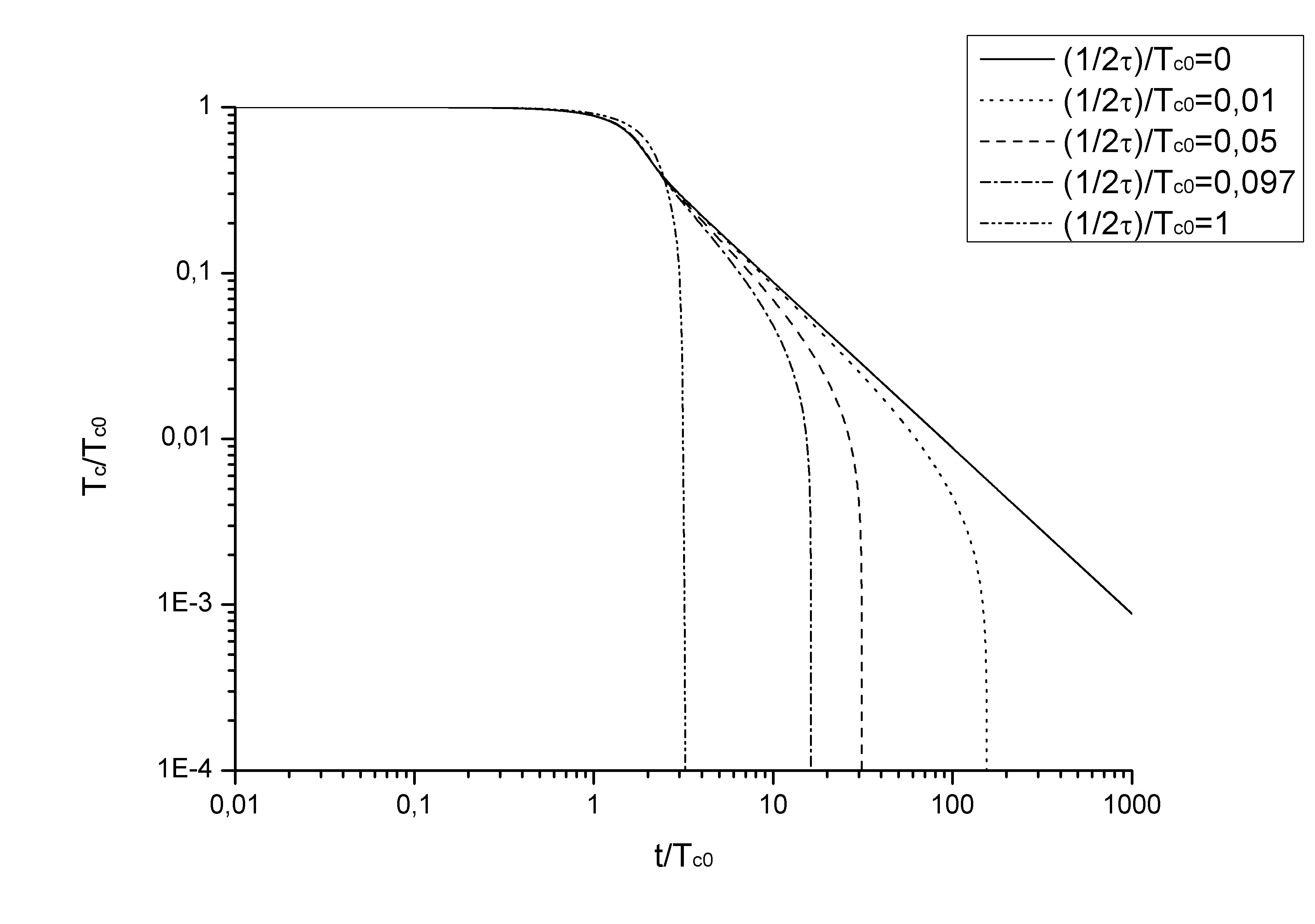}
\end{center}
\caption{Graph of $T_{c}/T_{c0}$ as a function of $t/T_{c0}$. The clean case
$\left(  1/2\tau=0\right)  $ is presented by the solid line. The impurity are
plotted with repectively $\left(  1/2\tau\right)  /T_{c0}%
=0.01(~0.05,~0.097,~1$ in dotted (respectively dashed, dashed-dotted and
dashed-dottted-dotted) line. We see that the impurities enhance the
superconducting transition temperature for weak interlayer coupling. On the
other hand superconducting critical temperature decreases quickly in the
presence of impurities at strong interlayer coupling.}%
\label{fig_impurete_Tc_fct_det}%
\end{figure}

We start with the analysis of the influence of the impurities on the
superconducting critical temperature. Then $\left(  \ref{f_general_SN}\right)
$ writes
\begin{equation}
f_{0,0}^{+}=\tfrac{\Delta^{\ast}\left(  t^{2}+2\left(  \omega+\frac{1}{2\tau
}\right)  ^{2}\right)  }{\left(  2\omega+\frac{1}{2\tau}\right)  t^{2}%
+2\omega\left(  \omega+\frac{1}{2\tau}\right)  ^{2}}. \label{f001}%
\end{equation}
in accordance with the model developed in \cite{Bulaevskii1}.The self
consistency equation $\left(  \ref{Self_consistency_1}\right)  $\ in this case
is written as%
\begin{equation}
\ln\left(  \tfrac{T_{c}}{T_{c0}}\right)  =2\pi T_{c}\sum_{\omega=0}^{\infty
}\left(  \tfrac{\left(  t^{2}+2\left(  \omega+\frac{1}{2\tau}\right)
^{2}\right)  }{\left(  2\omega+\frac{1}{2\tau}\right)  t^{2}+2\omega\left(
\omega+\frac{1}{2\tau}\right)  ^{2}}-\tfrac{1}{\omega}\right)  .
\label{Self_consistency_1_impur}%
\end{equation}
In the case of weak proximity effect $\left(  t\ll T_{c0}\right)  $, the
decrease of the critical temperature is deduced from the equation $\left(
\ref{Self_consistency_1_impur}\right)  $and reads%
\[
\tfrac{T_{c}-T_{c0}}{T_{c0}}=\tfrac{\Delta T_{c}}{T_{c0}}=-\tfrac{1}{2}\left(
\tau t\right)  ^{2}\left(  \left(  \tfrac{1}{2\tau T_{c0}}\right)  \pi
+4\Psi\left(  \tfrac{1}{2}\right)  -4\Psi\left(  \tfrac{1}{2}+\tfrac{1}{4\tau
T_{c0}\pi}\right)  \right)  .
\]
In the clean limit $\left(  T_{c0}\gg t\gg\frac{1}{2\tau}\right)  $ the
superconducting critical temperature varies as $\frac{T_{c}}{T_{c0}}=1-\left(
\frac{7}{8}\frac{\zeta\left(  3\right)  }{\pi^{2}}-\frac{\pi}{192}\frac
{1}{\tau T_{c0}}\right)  \frac{t^{2}}{T_{c0}^{2}}$, and the impurity
scattering inside the N layer decreases the proximity effect. In the dirty
regime $\left(  T_{c0}\gg\frac{1}{2\tau}\gg t\right)  $ the superconducting
critical temperature varies as $\frac{T_{c}}{T_{c0}}=1-\frac{\pi}{2}\frac{\tau
t^{2}}{T_{c0}}$. The presence of impurities enhanced the superconducting state
and $T_{c}$ decreases slower than in the clean case (see figure
$\ref{fig_impurete_Tc_fct_det}$). In this case, the impurities decreases the
effective transfer coupling and then the proximity effect.

However at strong interlayer coupling $t\gg T_{c0}$ and
$1/2\tau\ll T_{c}$, the expression for the anomalous Green function $\left(
\ref{f001}\right)  $ becomes $f_{0,0}^{+}=\tfrac{\Delta^{\ast}\left(
t^{2}+2\omega^{2}\right)  }{2\omega\left(  t^{2}+\omega^{2}\right)  }\left(
1+\left(  \tfrac{t^{2}\left(  2\omega^{2}-t^{2}\right)  }{2\omega\left(
t^{2}+\omega^{2}\right)  \left(  t^{2}+2\omega^{2}\right)  }\right)  \left(
\tfrac{1}{2\tau}\right)  \right)  $ so the critical temperature varies as
$\tfrac{T_{c}}{T_{c0}}=\frac{\pi e^{-\gamma}}{2}\frac{T_{c0}}{t}\left(
1-\frac{1}{8}\frac{t}{\tau T_{c0}^{2}}\right)  $. This means that scattering
on impurities strongly decreases $T_{c}$ for high interlayer coupling as seen
in the figure $\ref{fig_impurete_Tc_fct_det}$. In the regime $t\gg T_{c0}$,
the mixing between the superconducting state in the S layer and the normal
state in the N layer is very strong. The bilayer draws near the regime
$\widetilde{\lambda}\longrightarrow\lambda/2$ where the S-N bilayer can be
considered as a single S layer with an effective coupling constant
$\widetilde{\lambda}<\lambda$. Note that $T_{c}$ depends on the impurities
contrary to the Anderson theorem prediction which is not astonishing because
the system is non uniform.

\subsection{Effect of the impurities on the phase diagram.}

In this section, we study the influence of the impurities on the $\left(
h,T\right)  $\ and $\left(  h,t\right)  $ phase diagram of the S-N bilayer. In
the presence of impurities, the modulated phase disappears \ and $h^{FFLO}$
decreases to $h^{I}$ \cite{Aslamazov}$^{,}$\cite{Bulaevskii3}. When the normal
phase is overcooled then the critical magnetic field decreases from $h^{I}$ to
$h_{0}$ \cite{D St James et Sarma}. For simplicity in the whole paper, we will
focus on the second order transition critical field of the S-N bilayer, taking
in mind that if the transition is of order the first order the calculated
field corresponds to the overcooling field and the critical region of
superconductivity phase existence may be somewhat larger. Consequently, we
study the influence of the impurities in the homogeneous case $\left(
q=0\right)  $. In this case, the anomalous Green function is the same as
$\left(  \ref{f001}\right)  $ with the substitution $\omega\longrightarrow
\omega+ih$ and can be written as
\begin{equation}
f_{0,0}^{+}=\tfrac{\Delta_{0}^{\ast}\left[  2\left(  \omega+ih+\frac{1}{2\tau
}\right)  ^{2}+t^{2}\right]  }{\left(  2\left(  \omega+ih\right)  +\frac
{1}{2\tau}\right)  t^{2}+2\left(  \omega+ih\right)  \left(  \omega+ih+\frac
{1}{2\tau}\right)  ^{2}} \label{f002}%
\end{equation}

\subsubsection{$\left(  h,t\right)  $ phase diagram at zero temperature}

\begin{figure}[ptb]
\begin{center}
\includegraphics[scale=0.3,angle=0]{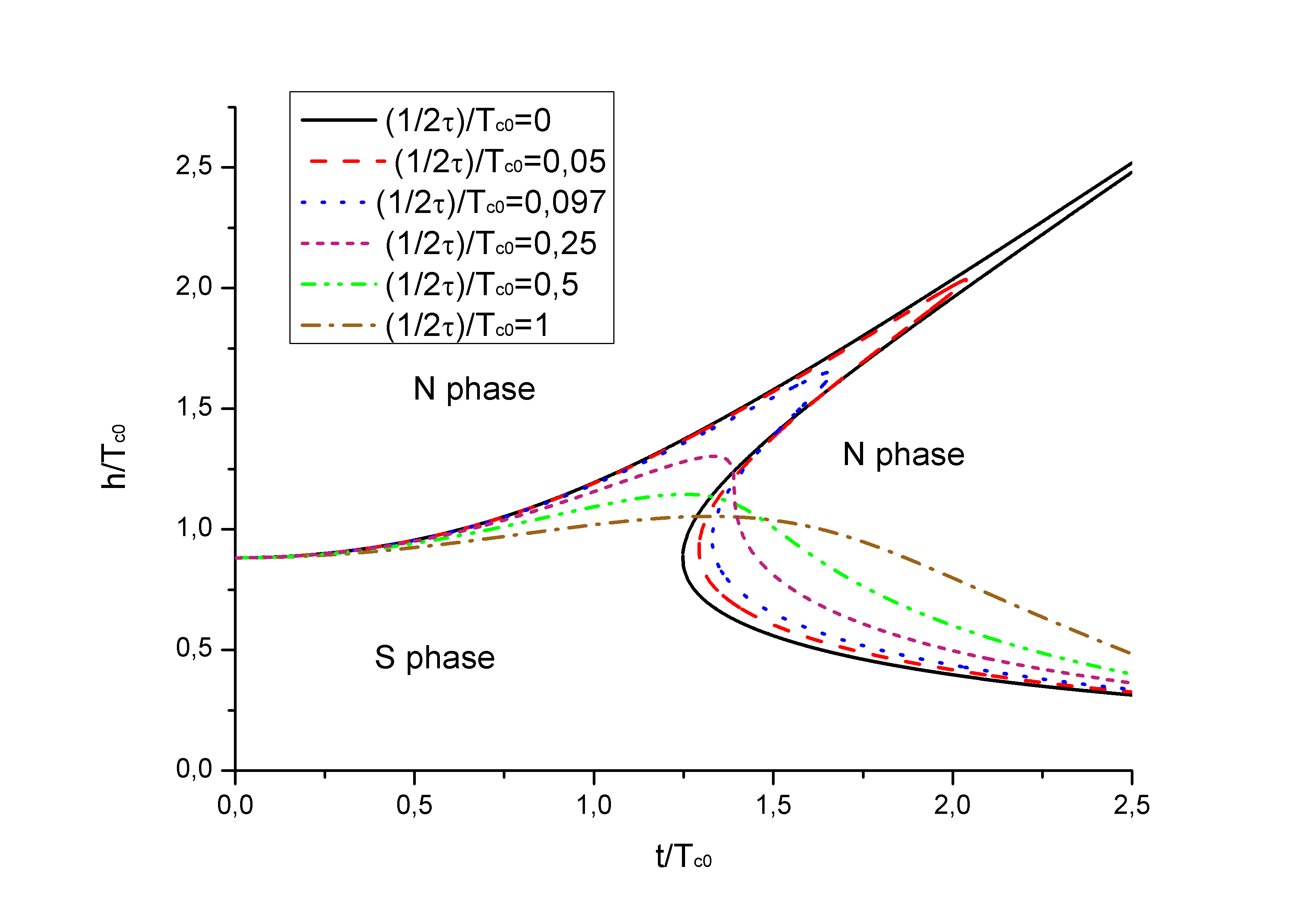}
\end{center}
\caption{$\left(  h/T_{c0},t/T_{c0}\right)  $ diagram for the S-N bilayer in
the clean case $\left(  \tau\rightarrow\infty\right)  $ (solid line) and with
respectively $\left(  1/2\tau\right)  /T_{c0}=0.05\left(  0.097,1\right)  $ in
dashed (respectively dotted, dashed-dotted) line. }%
\label{fig_hfcttSNC}%
\end{figure}

The impurities change the form of the $\left(  h,t\right)  $ phase diagram at
$T=0K$ as shown presented in the figure $\ref{fig_hfcttSNC}$. The $\left(
h,t\right)  $ phase diagram has been calculated numerically. The critical
interlayer coupling $t_{c}$ increases with the impurities diffusion potential
$1/2\tau$. The maximal values of $h_{c1}$ and $h_{c2}$ decreases with the
impurities diffusion potential contrary to $h_{c3}$. The variations of
$h_{c1}$, $h_{c2}$ and $h_{c3}$ reveals that the superconducting phase in the
S layer is enhanced by the presence of the impurities whereas the field
induced superconducting phase is destroyed by the impurities.

\subsubsection{$\left(  h,T\right)  $ phase diagram}

\begin{figure}[ptb]
\begin{center}
\includegraphics[scale=0.3,angle=0]{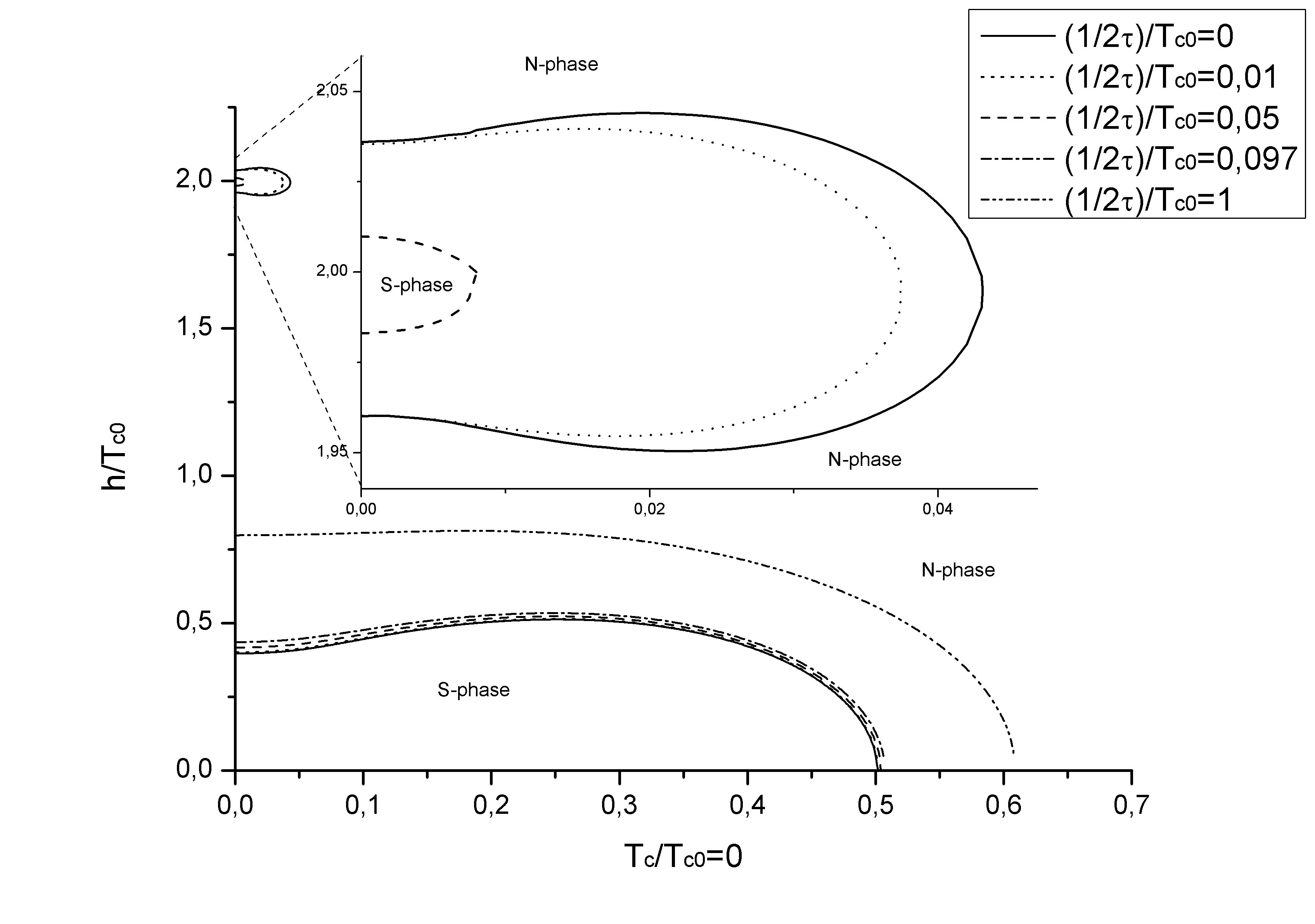}
\end{center}
\caption{$\left(  h/T_{c0},T_{c}/T_{c0}\right)  $phase transition diagram for
the S-N bilayer calculated for $t=2T_{c0}$ with the second order transition
line in the clean case $\left(  \left(  1/2\tau\right)  /T_{c0}=0\right)  $
(solid line) and with respectively $\left(  1/2\tau\right)  /T_{c0}%
=0.01\left(  0.05,~0.097,~1\right)  $ in dotted (respectively dashed,
dashed-dotted, dashed-dotted-dotted) line. The inset presents a zoom of the
superconducting re-entrance phase around $h\simeq t\simeq2T_{c0}$.}%
\label{fig_impureteSN_Tc_fct_det}%
\end{figure}

The $\left(  h,T\right)  $ phase diagram as been calculated numerically. The
reentrance phase is strongly influenced by the presence of the impurities as
seen on the figure $\ref{fig_impureteSN_Tc_fct_det}$. The maximal critical
temperature under which the field induced phase exists, decreases with the
impurity scattering potential. Moreover the upper and lower critical fields of
the re-entrant superconducting phase ($h_{c1}$ and $h_{c2}$) become closer
with the increase of impurities diffusion potential as seen in the last part.
In the case $t=2T_{c0}$, the reentrance phase totally disappears for an
impurity diffusion potential $1/2\tau$ upper than $0.097T_{c0}$. In the figure
$\ref{fig_impureteSN_Tc_fct_det}$, the usual superconducting phase is also
influenced by the presence of impurities. The critical magnetic field at zero
temperature $h_{c3}$ \ and the critical temperature at zero magnetic field
$T_{c}$ increase with the impurities diffusion potential. The effective
interlayer coupling decreases in the presence of impurities then the usual
superconductivity in the S layer is enhanced. 

\section{EFFECT OF THE IMPURITIES ON THE S-S BILAYER}

In this section, we study the S-S bilayer considering the FFLO modulation and
the impurities. As predicted in \cite{Andreev1} for ferromagnet superconductor
multilayered systems, a $\pi$ state may appear in S-S bilayer under magnetic
field. Using the same model as developed in the section II, the S-S bilayer is
described by the following equations%
\begin{equation}
\left(
\begin{array}
[c]{cccc}%
\widetilde{\omega}-i\mathbf{v}_{F}.\mathbf{q} & -i\frac{t}{2} & 0 & i\frac
{t}{2}\\
-i\frac{t}{2} & \widetilde{\omega}-i\mathbf{v}_{F}.\mathbf{q} & i\frac{t}{2} &
0\\
0 & i\frac{t}{2} & \widetilde{\omega}-i\mathbf{v}_{F}.\mathbf{q} & -i\frac
{t}{2}\\
i\frac{t}{2} & 0 & -i\frac{t}{2} & \widetilde{\omega}-i\mathbf{v}%
_{F}.\mathbf{q}%
\end{array}
\right)  .\left(
\begin{array}
[c]{c}%
f_{0,0}^{+}\\
f_{1,0}^{+}\\
f_{1,1}^{+}\\
f_{0,1}^{+}%
\end{array}
\right)  =\left(
\begin{array}
[c]{c}%
\Delta_{0}^{\ast}+\frac{\left\langle f_{0,0}^{+}\left(  \omega,\mathbf{q}%
\right)  \right\rangle _{\phi}}{2\tau}\\
0\\
\Delta_{1}^{\ast}+\frac{\left\langle f_{1,1}^{+}\left(  \omega,\mathbf{q}%
\right)  \right\rangle _{\phi}}{2\tau}\\
0
\end{array}
\right)  \label{equation_Eilenberger_SS1}%
\end{equation}

where $\Delta_{1}^{\ast}$ is the superconducting gap in the S layer indexed
$j=1$.

In the $\pi-$phase, $\Delta_{0}^{\ast}=-\Delta_{1}^{\ast}$, the solution of
the system $\left(  \ref{equation_Eilenberger_SS1}\right)  $ is%
\[
f_{0,0}^{+}={\tfrac{\Delta_{{0}}^{\ast}}{2\,\left(  1-\frac{1}{2\tau}%
\frac{\left(  \Omega_{2}+\Omega_{1}\right)  }{2\,\Omega_{1}\Omega_{2}}\right)
}}\left(  \frac{1}{\omega_{1}}+\frac{1}{\omega_{2}}\right)
\]
and the averaged solution on the $\phi$ angle
\[
\left\langle f_{0,0}^{+}\right\rangle _{\phi}={\tfrac{\left(  \Omega
_{1}+\Omega_{2}\right)  \Delta_{{0}}^{\ast}}{2\,\Omega_{1}\Omega_{2}\left(
1-\frac{1}{2\tau}\left(  \Omega_{2}+\Omega_{1}\right)  \right)  }}%
\]

In the clean limit $\left(  \tau\longrightarrow\infty\right)  $, at zero
temperature, the $\pi$ superconducting phase appears above the critical
magnetic field $h_{low}=t-\Delta_{0}^{2}/8t$ and below $h_{up}=t+\Delta
_{0}^{2}/8t$ \ in the limit $t\gg\Delta_{0}$. As predicted in \cite{Buzdin2},
the modulated FFLO state appears at low temperature and maximize the critical
magnetic field. Hence, with the FFLO state, the critical magnetic fields are
$h_{low,up}=t\mp\Delta_{0}^{2}/4t$ in the limit $t\gg\Delta_{0}$. The
re-entrance superconducting phase is enhanced at low temperature by FFLO
modulations. The presence of impurities in the system may destroy the FFLO
state and the re-entrance phase. The FFLO transition should meet quickly the
first order transition line. Consequently, we will study the influence of the
impurities in the homogeneous case where $q=0$.

\subsection{$\left(  h,t\right)  $ phase diagram}

\begin{figure}[ptb]
\begin{center}
\includegraphics[scale=0.3,angle=0]{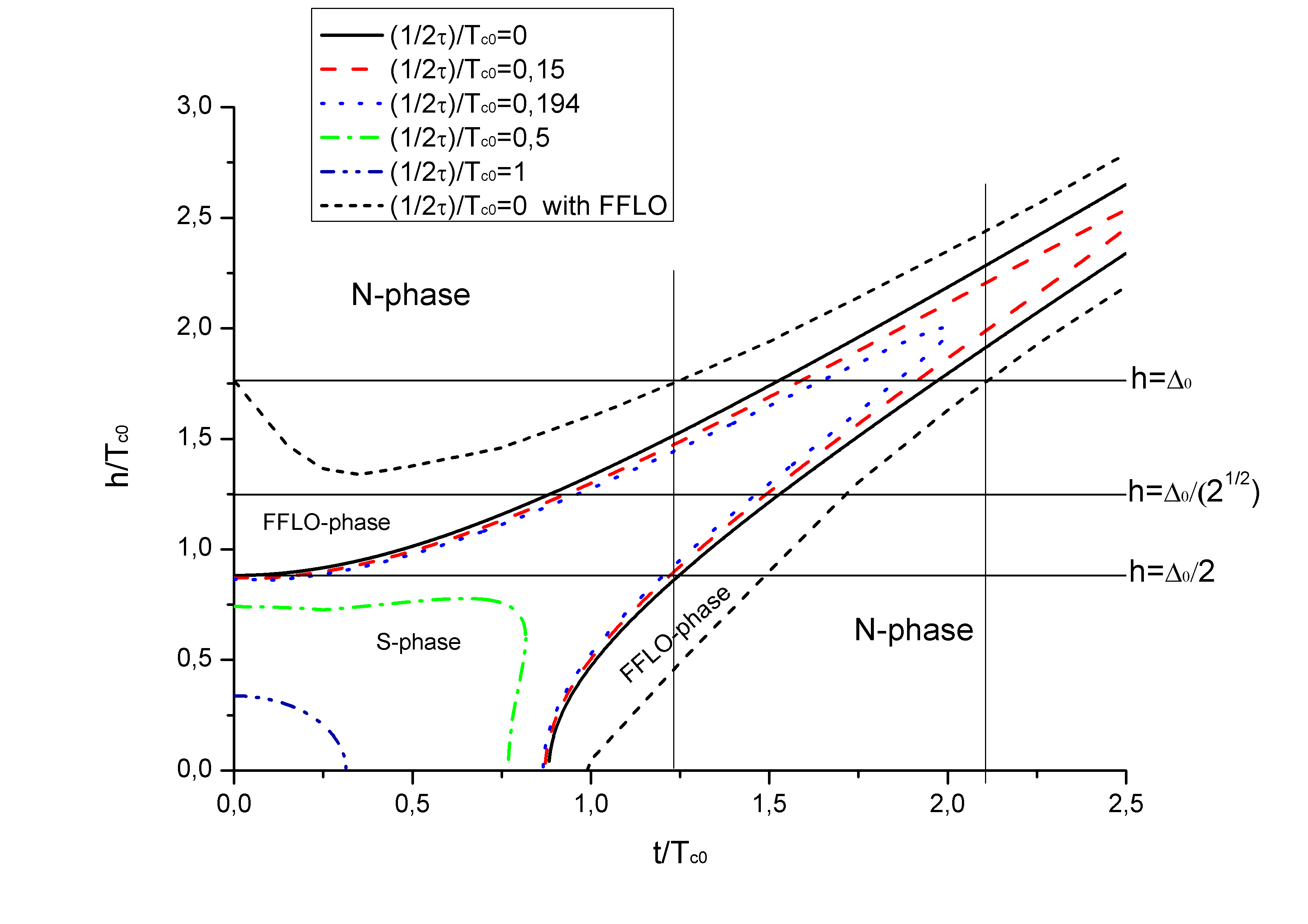}
\end{center}
\caption{$\left(  h/T_{c0},t/T_{c0}\right)  $ diagram for the S-S bilayer in
the clean case $\left(  \left(  1/2\tau\right)  /T_{c0}=0\right)  $ (solid
line) and with respectively $\left(  1/2\tau\right)  /T_{c0}=0.15\left(
0.194,~0.1,~0.5,1\right)  $ in dashed (respectively dotted,dashed-dotted and
dashed-dotted-dotted) line. The lines $h=\Delta_{0}/2$, $h=\Delta_{0}/\sqrt
{2}$ and $h=\Delta_{0}$ present respectively the critical magnetic field for a
second order superconducting phase transition, the critical magnetic field for
a first order phase transition and the FFLO paramagnetic limit for a single S
layer. The close-dashed line is the FFLO paramgnetic limit in the S-S bilayer
in the clean limit.}%
\label{Fig_hfcttSS}%
\end{figure}

At $T=0K$ \ in the $\pi$-state without FFLO modulation, the self consistency
equation $\left(  \ref{Self_consistency_2}\right)  $ becomes :%
\begin{equation}
\left(  4\left(  \tfrac{1}{2\tau}\right)  ^{2}+\left(  2h+\sqrt{\left\vert
-4t^{2}+\left(  \tfrac{1}{2\tau}\right)  ^{2}\right\vert }\right)
^{2}\right)  \left(  4\left(  \tfrac{1}{2\tau}\right)  ^{2}+\left(
2h-\sqrt{\left\vert -4t^{2}+\left(  \tfrac{1}{2\tau}\right)  ^{2}\right\vert
}\right)  ^{2}\right)  =16h_{0}^{2}. \label{selfSSsansFFLO}%
\end{equation}
The solutions of $\left(  \ref{selfSSsansFFLO}\right)  $ are $h_{up,low}%
^{imp}=\frac{1}{2}\sqrt{\left\vert -4t^{2}+\left(  \tfrac{1}{2\tau}\right)
^{2}\right\vert -4\left(  \tfrac{1}{2\tau}\right)  ^{2}\pm4\sqrt{-\left(
\tfrac{1}{2\tau}\right)  ^{2}\left\vert -4t^{2}+\left(  \tfrac{1}{2\tau
}\right)  ^{2}\right\vert +h_{0}^{4}}}$ where $h_{up,low}^{imp}$ are the
critical magnetic field of the S-S bilayer in the presence of impurities (see
figure $\ref{Fig_hfcttSS}$ ). The field induced superconducting state is
destroyed in presence of impurities and cannot be observed if $h_{low}%
^{imp}=h_{up}^{imp}$. We define a critical impurity diffusion time $\tau
_{c}=1/2\left(  \sqrt{2t^{2}-\sqrt{4t^{4}-h_{0}^{4}}}\right)  $ below which
the re-entrance phase totally disappears. In the case where $t=2T_{c0}$ and
$h_{0}=0.882T_{c0}$ then $\left(  \frac{1}{2\tau}\right)  _{c}\simeq
0.194T_{c0}$.

The critical magnetic field in the presence of FFLO modulation is
plotted in the figure $\ref{Fig_hfcttSS}$in the clean limit. The critical
magnetic field in the presence of FFLO modulations is the upper limit of the
critical magnetic field.

We can see that $h_{up}^{FFLO}$, the upper critical field in presence of FFLO
modulations, cross the line $h=\Delta_{0}$ for $t\simeq1.25T_{c0}$. It means
that the usual superconducting phase is deformed only for $t>1.25T_{c0}$ at
$T=0K$. Then the field induced superconducting phase becomes observable. The
field induced superconducting phase become totally observable when
$h_{low}^{FFLO}$, the lower critical field in presence of FFLO modulations,
cross the line $h=\Delta_{0}$ for $t\simeq2.1T_{c0}$.

In the uniform case, we would have to consider the first order transition
line. For the S-S bilayer, the first order transition line is between the
second order and the FFLO transition line. The reentrance phase would appear
when $h_{up}^{I}$, the upper critical field for a first order phase transition
is above $h^{I}=\Delta_{0}/\sqrt{2}$ and would be distinguishable if the lower
critical field in the case of first order phase transition is higher than $h$.

\subsection{$\left(  h,T\right)  $ phase diagram}

In the $\pi$-state, the Cooper pairs are formed by two electrons in the
different layer. The standard superconducting state is only due to the
$0-$phase and then is not influenced by the impurities as predicted by the
Anderson theorem. The lower and upper critical lines merge at field $h=t$\ and
temperature $T_{M}=\pi e^{-C}T_{c0}^{2}/\left(  4t\right)  $ in the limit
$t\gg T_{c0}$. The field induced $\pi$ superconductivity is confined to
temperature lower than $T_{M}$.

\begin{figure}[ptb]
\begin{center}
\includegraphics[scale=0.3,angle=0]{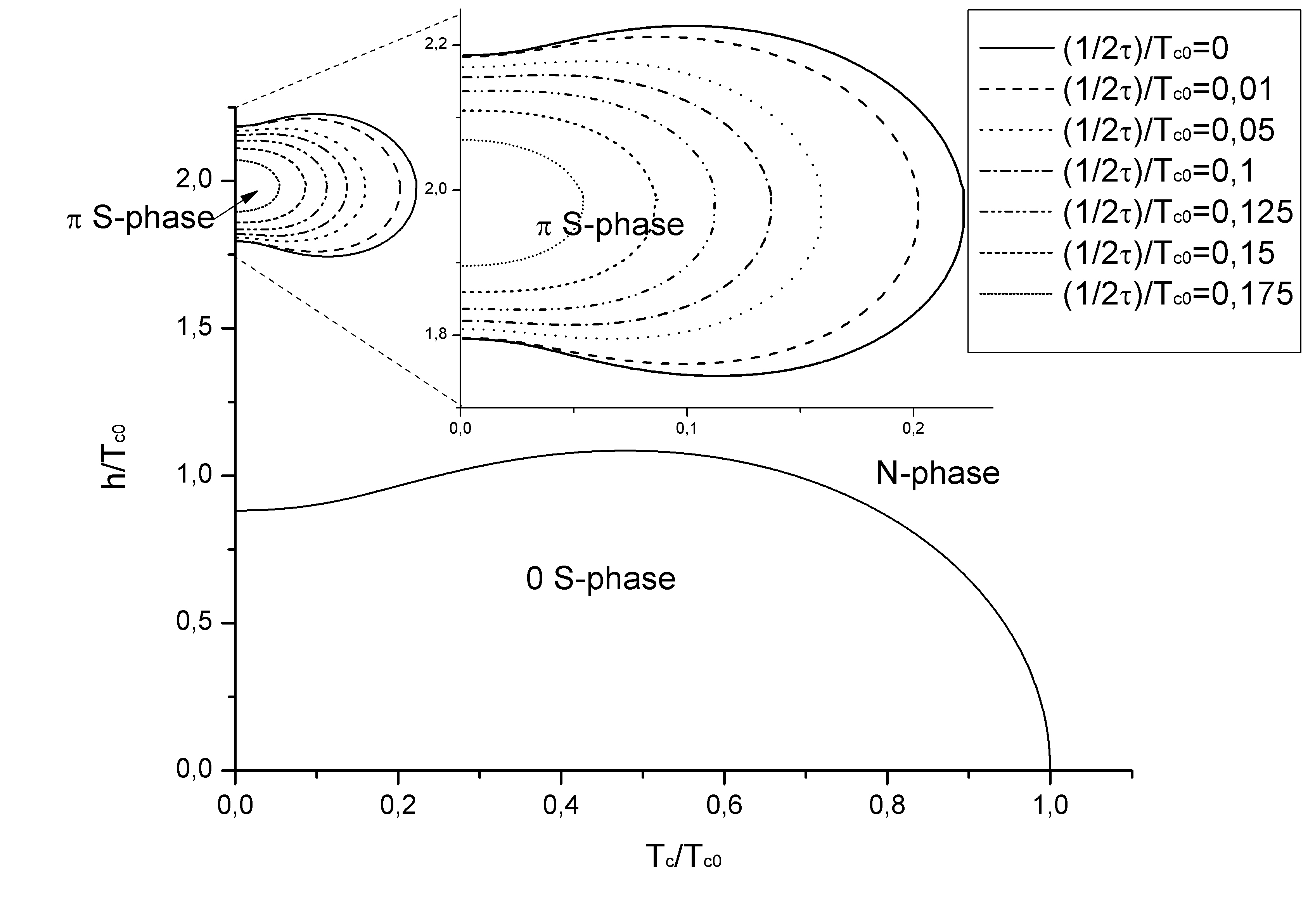}
\end{center}
\caption{$\left(  h/T_{c0},T_{c}/T_{c0}\right)  $phase transition diagram for
the S-S bilayer calculated for $t=2T_{c0}$ with the second order transition
line in the clean case $\left(  \left(  1/2\tau\right)  /T_{c0}=0\right)  $
(solid line) and with respectively $\left(  1/2\tau\right)  /T_{c0}%
=0.01\left(  0.05,~0.1,~0.125,~0.15,~0.175\right)  $ in dotted (respectively
dashed, dashed-dotted, dashed-dotted-dotted, close dashed, close dotted) line.
The inset presents a zoom of the superconducting re-entrance phase around
$h\simeq t\simeq2T_{c0}$.}%
\label{fig_impureteSSJ}%
\end{figure}

On the phase diagram, we see that the reentrance decreases as the impurity
self energy is increasing (see figure $\ref{fig_impureteSSJ}$). The reentrance
phase totally disappear for $\frac{1}{2\tau}\simeq0.194T_{c0}$ in the case
where $t=2T_{c0}$. The existence of first order transition line in the field
induced phase transition could influence these results. $h_{up,low}^{I}$ are
higher(smaller) than $h_{up,low}$. Consequently, the critical impurity
diffusion time $\tau_{c}$ should be higher than in the case of a second order
transition. 

\section{CONCLUSION}

To conclude, the proximity effect plays a crucial role in the S-N \ and S-S
\ bilayers. The superconducting critical temperature and the critical magnetic
field at zero temperature in the S-N and the S-S bilayers depends directly on
the interlayer coupling. We demonstrated that at low temperature, a magnetic
field induced superconducting phase appears at high in-plane magnetic field in
S-N bilayers. This field induced phase is originated from the compensation of
Zeeman effect energy splitting by the energy splitting between the bonding and
antibonding state electronic levels. This reentrance phase provides the
possibility to overcome the classical paramagnetic limit and the results of
our work give the hints for engineering layered superconducting material with
very high critical fields.

In S-S and S-N bilayers, the presence of impurities make the superconducting
field induced phase more difficult to observe. The impurities produce a
broadening of the different energy levels over an energy range $1/\tau$ which
prevents exact compensation. It is possible to define a critical mean free
pass time over which the re-entrance phase cannot survive. In the S-N and S-S
bilayer, the critical mean free pass time $\tau_{c}$ only depends on the
interlayer coupling. In S-S bilayer, in the case $t\simeq\Delta_{0}$ then
$\tau_{c}^{-1}\simeq0.25\Delta_{0}$ above which there is no possibility to
observe field induced phase. From an experimental point of view, it could be
possible with the molecular beam epitaxy techniques to provide a sufficiently
large mean free path \ to realize the condition of field-induced phase observation.

Although we have only treated the Zeeman effect as cooper pair breaking
effect, we have to discuss on the orbital pair breaking effect. In the case of
multilayered system under in-plane magnetic field, the condition for
neglecting the orbital effect is given by $tH\xi_{0}d/\Phi_{0}<\Delta_{0}$,
where $\xi_{0}$ is the in-plane coherence length and $\Phi_{0}=h/2e$ the
superconducting quantum of magnetic flux. In the case $t\simeq\Delta_{0}$we
obtain that $H$ must be lower than $H_{orb}\simeq\Phi_{0}/\left(  \xi
_{0}d\right)  $. The typical values $d\simeq10\mathring{A}$ , $\xi_{0}%
\simeq100\mathring{A}$ the corresponding field is extremely large
$H_{orb}\simeq200T$ and not restrictive at all as the maximal currently
attainable permanent magnetic field are $60T$. The orbital effect becomes
important in layered system in the case $t\gg\Delta_{0}$ when the Pauli limit
may be exceeded many times. However in \cite{Ledge}, it was demonstrated that
the orbital pair breaking in layered superconductors are switched off in the
high field regime and the superconductivity is restored. We may expect that
similar situation should be realized in S-N and S-S bilayer.

\begin{acknowledgments}
The authors would like to thanks S. Burdin an V.H. Dao for help and useful
discussions. This work was supported, in part, by French ANR "SINUS" program.
\end{acknowledgments}

\end{document}